**Title:** A beta-binomial model respecting randomization and its comparison to the standard beta-binomial model that ignores randomization for the meta-analysis of rare events

**Authors**

Tim Mathes (corresponding author)[1]

Tim.mathes@med.uni-goettingen.de

Maxi Schulz[1]

Maxi.schulz@med.uni-goettingen.de

Oliver Kuss[2,3]

Oliver.kuss@ddz.de

1. Department of Medical Statistics, University Medical Center Göttingen, Germany
2. German Diabetes Center, Institute for Biometrics and Epidemiology, Germany
3. Centre for Health and Society, Medical Faculty and University Hospital Düsseldorf, Heinrich Heine University Düsseldorf, Germany

## Abstract

***Background***

One of the suggested models for meta-analysis with rare events is the beta-binomial model (BBM). The main advantage of this model compared to inverse-variance models, is that it uses information from zero cells without needing a continuity correction. A disadvantage of the "standard" BBM is that it ignores randomization.

Here we introduce a BBM that respects randomization.

***Methods***

The main idea to preserve randomization is using a "common-beta" BBM. We illustrate that randomization is preserved by conditioning on the total sum of counts in a study's four-fold table when estimating the model parameters. We perform a simulation study reflecting real-world meta-analyses to compare the models. In addition, we explore in which situations ignoring randomization could be problematic.

***Results***

The BBM that respects randomization performs well in the simulation study that mirrors real meta-analyses in Cochrane and non-Cochrane reviews, respectively. Ignoring randomization appears to be problematic in situations with very different sample sizes of the studies included in the meta-analysis. However, the BBM ignoring randomization tended to perform better when heterogeneity was high.

***Conclusion***

The results show that using the standard BBM, which ignores randomization is usually not biased when the randomization is balanced and the size of studies included in the meta-analysis is not very different. However, as possible ecological bias due to ignoring randomization is an inherent disadvantage of the model and the BBM that respects randomization shows very similar results in the simulation study, it may be generally preferred.

**Background**

Rare events are a common situation when performing meta-analyses [1]. In systematic reviews on harms, rare events are the rule rather than the exception [2]. In the case of rare events, primary studies are usually underpowered and meta-analyses are an attractive way to increase power. However, meta-analyses of rare events are challenging, particularly when studies with zero events are involved. In the past, the standard approach for this situation was using either methods that can deal with these situations (e.g. Peto method, Mantel-Haenszel method) or using continuity correction to avoid cells with zero observations. The drawback of the first method is that these are fixed-effect models and thus are not adequate when heterogeneity is present. The drawback of the second is that artificially adding numbers to study cells can lead to inflated standard errors and biased results [3].

Although many methods exist for rare event meta-analyses, including choices for pooling methods, heterogeneity variance estimators, and ways to correct data with zero events, application in practice is often inconvenient because their performance depends on their combination and the data at hand [3]. Furthermore, in the case of rare events random-effects meta-analyses often fall back to a fixed-effect model [4].

For binary outcomes, four-fold tables for each study can be reconstructed. For this reason, it is possible to use regression-based models for binary outcomes for meta-analyses. Several regression-based meta-analysis methods including Bayesian approaches have been suggested previously [5-8]. The main advantage of the proposed models is that they use information from zero cells without the need for continuity corrections.

One of the suggested models is the beta-binomial model often used for modelling overdispersed binary data (e.g. in economics). As the baseline success probability is considered a random effect that follows a beta-distribution it is a “true“ random-effects model that will never collapse to a common-effect model. It is worth pointing out that the beta-binomial model differs from the standard random-effects models as understood in the meta-analytic context, where the treatment effect is assumed to be random. When using beta-binomial models the treatment effect is not directly modelled as a random effect but varies only implicitly. Several simulation studies have shown good statistical properties of the beta-binomial model for meta-analyses of rare events [7, 9-12].

A disadvantage of the previously introduced beta-binomial model is that it ignores randomization in the sense that not the contrasts (differences between treatment and control) within the study but the control arms are pooled first across studies and then linked together to arrive at the final estimate of the treatment effect. This is in contrast to the standard inverse variance meta-analysis methods, where the contrast is estimated first for each study and the results are pooled thereafter. In the network-meta-analyses context, the different approaches are known as arm-based and contrast-based [13]. Although the results of previous simulations suggest that pooling arms first may not be problematic in practice, concerns remain because it cannot be excluded that ignoring randomization may be problematic in some specific situations [13].

In a previous work, a common-beta beta-binomial model that respects randomization was assessed in a simulation study for the situation of very few studies [10]. In this study this beta-binomial model showed comparable results to the standard (common-rho) beta-binomial model [10].

Here we consider the beta-binomial model that respects randomization in more depth. Specifically, we describe the characteristics that ensure respecting the randomization and illustrate the differences of the models. In addition, we compare the models in a simulation study. Finally, we explore in which situations ignoring randomization could be problematic by comparing the beta-binomial models in different situations.

The comparison of BBMs that ignore randomization to models that respect randomization provides insights when using models that ignore randomization may be problematic and thus can support choosing the suitable model for the specific data at hand.

## Methods

### *Preliminaries*

In the following we consider meta-analyses including $K$ studies to compare two treatments (T for treatment and C for control) using a binary outcome. For each study $k$ ($k = 1, \ldots, K$), $n_{kT}$ and $n_{kC}$ denote the sample size for the treatment and control, $y_{kT}$ and $y_{kC}$ the number of events and $z_{kC}$ the observations in the control arm.

### *Models*

*Standard (common-rho) beta-binomial models*

In the standard beta-binomial model (BBST) it is assumed that observed numbers of events in the control arm $y_{kC}$ ($k = 1, \ldots, K$) follow a binomial distribution $Bin(\pi_{kC}, n_{kC})$. The event probability $\pi_{kC}$ follows a beta distribution with parameters $\alpha_C$ and $\beta_C$ across all control arms. The expected value and variance of $\pi_{kC}$ are:

$$E(\pi_{kC}) = \mu_C$$

and

$$Var(\pi_{kC}) = \mu_C \times (1 - \mu_C) \times \nu_C / (1 + \nu_C)$$

with

$$\mu_C = \alpha_C / (\alpha_C + \beta_C)$$

and variance

$$\nu_C = 1 / (\alpha_C + \beta_C)$$

Thus, the number of events $y_{kC}$ is beta-binomially distributed with expected value

$$E(y_{kC}) = n_{kC} \times \mu_C$$

and variance

$$Var(y_{kC}) = n_{kC} \times \mu_C \times (1 - \mu_C) \times [1 + (n_{kC} - 1) \times \nu_C / (1 + \nu_C)].$$

The correlation between two individual binary observations $\rho_C = corr(z_{kCj_1}, z_{kCj_2})$ in the control arm of study $k$ ($k = 1, \ldots, K; j_1, j_2 = 1, \ldots, n_{kC}; j_1 \neq j_2$) is

$$\rho_C = 1 / (\alpha_C + \beta_C + 1),$$

and analogously for $\rho_T$ and two binary events in the treatment arm of a study.

Furthermore, in the common-rho beta-binomial model it is assumed that $\rho_C = \rho_T = \rho$, and consequently $\alpha_C + \beta_C = \alpha_T + \beta_T$.

The treatment effect $\theta_T = g^{-1}(\mu_T)/g^{-1}(\mu_C)$ is modelled via the link function (e.g., the logit)

$$g(\mu_i) = b_0 + b_T \times i$$

where $b_0$ denotes the logit of the risk of an event in the control arm, and analogously for other link functions like the log or the identity link.

When using the common-rho beta-binomial model the control group arm is pooled first and then the treatment effect is estimated by the link function. The beta distribution in the intervention group then results from the beta distribution in the control group, the link function and the common-rho assumption. This arm-based pooling ignores randomization, that is the information to which study a treatment group belongs is lost (i.e., treated individuals are not only directly compared to randomized controls of the same study).

*Common-beta beta-binomial models ignoring or respecting randomization*

As given above, for the BBST it is assumed that the intra-class correlation $\rho$ is equal in the two treatment arms implying that $\alpha_C + \beta_C = \alpha_T + \beta_T$. Another possibility is to assume that $\beta$ is equal in both groups leading to a common-beta beta-binomial model (BBCB), so that $\beta_C = \beta_T = \beta$ but the alphas ($\alpha_C$ and $\alpha_T$) differ between the treatment groups. It has been shown that the fixed-effects negative binomial regression (FE-NegBin) model for panel count data is equivalent to this common-beta BBM when estimating the parameters by conditional maximum likelihood (ML) [14]. These models are commonly used in econometrics [15]. Each study (in econometrical terms: "panel") contributes four cells, that is $y_{kT}$, $n_{kT} - y_{kT}$, $y_{kC}$, and $n_{kC} - y_{kC}$ (i.e., a fourfold table) for estimating the model.

In our previous publication, we introduced the common-beta beta-binomial model and suggested conditioning separately on the counts in the treatment and control groups of each study [14, 16]. However, conditioning on two counts, the number of patients in the treatment ($n_{kT}$) and control group ($n_{kC}$), for each study separately likewise ignores randomization. Therefore, in the following, we refer to this model as the "BBCBignor" model.

Alternatively, the ML estimation can be conditioned on the sum of all four counts in each study by summing up the counts from the treatment and control groups ($n_k$). Proceeding this way links the treatment and control group of a single study directly and thus respects randomization. We refer to this model as the "BBCBrespect" in what follows.

BBCBrespect and BBCBignor have similar likelihood functions (see table 1), the subtle, but decisive difference being that in the first six terms the BBCBignor log-likelihood sums over the terms of treatment and control group *separately*, whereas the BBCBrespect log-likelihood sums over the parallel three terms of treatment and control group *simultaneously*.

Figure 1 illustrates the different summarization levels when estimating the models and the consequences of respecting or ignoring randomization for the estimation.

Both models can be implemented using estimation methods for count data models for panel data as used in economics. Here, we used PROC COUNTREG in SAS.

***Simulation***

*Simulation design*

The data were generated from an inverse-variance random-effects model. We aimed to replicate real meta-analyses as closely as possible. Therefore, we generated all values based on input parameters and distributional assumptions deduced from the Cochrane Database of Systematic Reviews and a previous meta-research study on systematic review characteristics [17, 18]. Table 2 shows the distributional assumptions and descriptions of the resulting data values.

We simulated two different scenarios with regard to the number of studies. Within each of these number of study scenarios, the number of studies was varied randomly. The generation of the mean and variance of the number of studies were informed by the numbers given by Turner et al. [18], and Page et al. [17], reflecting the number of studies included in Cochrane (CO), and non-Cochrane (NC) reviews, respectively. For the CO this is a median number of studies of 3, and for the NC scenario a median number of 9 studies. Each of these scenarios was combined with a fixed baseline event probability in the control group of 0.01, 0.02, and 0.05. Finally, we simulated the data under the null/alternative hypothesis (H0/H1) of a treatment effect resulting in 12 different scenarios in total. For each scenario, we simulated 10,000 meta-analyses. The number of zero-event studies was not explicitly varied but naturally emerged as a result of the very low event probabilities. The average proportion of meta-analyses including at least one zero event study ranged from 0.70 (Cochrane scenario, p= 0.05, H0) to 0.99 (Non-Cochrane scenario, p=0.01, H1). Table 3 shows the average proportion of single zero studies for each arm and double zero studies within the simulated meta-analysis for each scenario. The proportion of single zero studies ranged from 0.16 to 0.62 and the proportion of double zero studies from 0.06 to 0.36. For each scenario we simulated 10.000 meta-analyses.

For all beta-binomial models we calculated 95%-CIs using the t-distribution for the treatment effect $b_T$

$$\hat{b}_{T(BB)} \pm t_{df;\,0.975} \times \hat{\sigma}_{BB}\ ,$$

where $\hat{\sigma}_{BB}$ the corresponding estimated standard error, and $t_{df;\,0.975}$ the 0.975 quantile of the t-distribution with $df = 2K - 2$ degrees of freedom. We used this number of degrees of freedom because previous simulation studies have shown that this is the best choice with regard to coverage probability [12].

*Performance measures*

To assess the performance of the methods we calculated the following measures:

- Number of converged simulation runs ($R$)
- Bias: Difference between the estimated and true effects $\hat{\theta}_r - \theta_r; r = 1, \dots, R$.
- Coverage probability: Proportion of converged simulation runs where the 95% CI included the true effect $\theta_r; r = 1, \dots, R$.
- Length of 95% CI: Difference ($CI_{U,r} - CI_{L,r}$) of upper ($CI_{U,r}$) and lower ($CI_{L,r}$) confidence limit of the 95% CI for $\theta_r; r = 1, \dots, R$.
- Power under $H_1$: Proportion of converged simulation runs under $H_1$ where the 95 % CI excluded the null effect.

We prepared Bland-Altman-Plots for the NC, p=0.2 scenario to compare BBST with BBCBignor, and BBCBignor with BBCBrespect to get insights into possible consequences of the common-rho versus the common-beta assumption, and the consequences of ignoring randomization, respectively.

In the main text, only the results for H1 are shown because the results of H0 and H1 were very similar.

We run the simulation in SAS/STAT® software Version 9.4 (SAS Institute Inc., Cary, NC, USA). All plots were prepared with R-Studio version 2024.04.2+764. The simulation code including the calculation of the performance measures and code for preparing figures is available in the supporting information (supplement 1).

To compare the beta-binomial models to common meta-analyses methods, we calculated an inverse-variance random effects meta-analysis (REM) as a reference, particularly to check if this model produces results closer to the BBCBrespect than to the BBCBignor and compare the models in situations where BBCBignor and BBCBrespect differ. We decided to use the Paule-Mandel (PM) heterogeneity variance estimator ($\tau^2$) and modified Hartung-Knapp (MHK) 95%-CIs as this is one of the preferred choices for inverse-variance meta-analysis methods [19]. The treatment effect of each study is given with

$$\hat{\theta}_k = \theta_k + \varepsilon_k \text{ with } \theta_k = \theta_{REM} + \delta_k,\ \delta_k \sim N(0, \tau^2) \text{ and } \varepsilon_k \sim N\left(0, \sigma_k^2\right).$$

And the overall effect with

$$\hat{\theta}_{REM} = \frac{\sum_{k=1}^{K} w_{k(REM)} \times \hat{\theta}_k}{\sum_{k=1}^{K} w_{k(REM)}},$$

The PM estimator is derived from the generalised Q statistic

$$Q(\tau_{PM}^2) = \sum_{k=1}^{K} w_{k(PM)} \times \left(\hat{\theta}_k - \hat{\theta}(\tau_{PM}^2)\right)^2$$

by setting $Q(\tau_{PM}^2)$ to its expected value $K-1$ with $w_{k(PM)} = 1/\left(\sigma_k^2 + \tau_{PM}^2\right)$ and $\hat{\theta}(\tau_{PM}^2) = \left(\sum_{k=1}^{K} w_{k(PM)} \times \hat{\theta}_k\right)/\left(\sum_{k=1}^{K} w_{k(PM)}\right)$ and solved iteratively.

The original Hartung-Knapp 95%-CI is given with

$$\hat{\theta}_{HK} \pm t_{K-1;\,0.975} \times \sqrt{q} \times \hat{\sigma}_{HK}$$

where $t_{K-1;\,0.975}$ is the 0.975 quantile of the t-distribution with $K-1$ degrees of freedom,

$$q = \frac{1}{K-1} \sum_{k=1}^{K} w_{k(HKSJ)} \times \left(\hat{\theta}_k - \hat{\theta}_{HKSJ}\right)^2$$

and

$$\hat{\sigma}_{HKSJ} = \sqrt{1\Big/\left(\sum_{k=1}^{K} w_{k(HKSJ)}\right)}.$$

As this confidence interval may become too short in very heterogenous data situations, we used the modified KH (MHKPM) confidence intervals where $q^* = \max\{1, q\}$ [20].

***Empirical example***

We illustrate the BB methods using an example data set from a Cochrane review including 60 studies (9,044 participants) assessing the effect of on-pump versus off-pump artery bypass grafting on stroke [21].

**Results of the simulation**

Figure 2 shows the number of converged runs under H1 (for H0 see supplement 2). The convergence of the BBST was around 95% in most situations. Convergence of BBCBignor and BBCBrespect were very similar ranging from about 88% in the CO, p=0.01 scenario to 95% in the NC, p=0.05 scenario. As for the MHKPM a continuity correction is applied, it converged in all runs.

Figure 3 shows the density plots for bias under H1 (for H0 see supplement 3). Bias was most pronounced in very rare event situations, which is equivalent to a higher proportion of zero-event studies. MHKPM showed the strongest and most frequently biased estimates. In the p=0.01 scenario, the BBST showed some extremely negative errors. For the common-beta beta-binomial models and the highest event probability scenarios (p=0.05), the bias was very similar and rather low. The distributions for bias were similar for the BB models, particularly the bias distributions of BBST and BBCBignor largely overlapped.

Figure 4 describes the coverage probability and figure 5 the length of the 95%-CIs under H1 (for H0 see supplement 4 and 5, respectively). In both scenarios, coverage probabilities deviate from the nominal 95%. While BBCBignor and BBCBrespect maintain or slightly exceed 95% coverage, with BBCBignor reaching about 0.97 at p = 0.05 in the CO scenario, BBST starts at 0.90 for p = 0.01 but increases to around 0.96 at p = 0.05. In the NC scenario, all models show the lowest coverage at p = 0.02 (around 0.92 for BBCBignor, 0.91 for BBCBrespect, and 0.93 for BBST). Coverage improves with higher event probabilities, reaching about 0.96 for BBCBignor, 0.94 for BBCBrespect, and 0.97 for BBST at p = 0.05. The lengths of the 95%-CIs of the BB-models were generally very similar. The length of the 95%-CIs decreased with an increasing number of studies and event probabilities. In the CO scenario coverage probability of MHKPM was above 0.97 for all event probabilities. Accordingly, MHKPM showed very wide 95%-CIs very often. In the NC scenario coverage probability of MHKPM was slightly below BBST and BBCBignor, this means MHKPM was more similar to BBCBrespect.

The results for power are presented in figure 6. In the CO scenario, the power was generally very low and highest for the BBST. In the NC scenario, the power was very similar for all BB-models. The power of MHKPM was very low.

The Bland-Altman-Plots (figures 7 and 8) illustrate the agreement of the logORs. In both comparisons, the difference in effect estimates increased with an increasing effect size. When looking at the cases where the logORs disagree it becomes apparent that most effect estimates originate from cases that would be considered as “not converged” for any model, as the logORs are unrealistically large (e.g. >100).

As can be seen in figure 7 the common-rho assumption tends to result in larger effects, or, in other words, the effect estimates from the BBST tend to be closer to the null as compared to the common-beta assumption.

Figure 8 shows that for the comparison of BBCBrespect and BBCBignor the difference between the logOR went in both directions. The x-shaped appearance means that effects tend to either agree or disagree depending on the effect size. Looking at the meta-analyses where BBCBrespect but not BBCBignor provides plausible values reveals that this were mostly situations in which the sample sizes and number of double zero studies varied widely between the included studies (e.g., three small studies with a sample size of about 100 and one large studies with a sample size of 1000). As MHKPM also showed plausible values in such situations (data not shown). Looking at the cases where BBCBignor provides more plausible values than BBCBrespect revealed that this were mainly meta-analysis with very high heterogeneity (e.g. tau-square >100). For illustration of such situations two example meta-analyses (one for different study sizes, and one with very high heterogeneity) from the simulation are given in table 4.

***Empirical example***

Figure 9 shows the effect estimates with 95%-CI for the example data set. The example analysis was in agreement with what can be expected from the simulation and the findings from the Bland-Altman-Plot. The logOR and 95%-CIs were very similar for the BBST and BBCBignor. The effect estimate of the BBCBrespect is slightly closer to the null and the confidence interval is slightly narrower. MHKPM was much closer to the null than all other models. Of note, when using MHKPM the 35 double zero studies are excluded from the analysis.

**Discussion**

In this work, we have introduced a beta-binomial model that respects randomization. This is achieved by, first, going from the standard common-rho beta-binomial model to a common-beta beta-binomial model which is equivalent to a panel count data (FE-Negbin) model and, second, by conditioning parameter estimation in this model on the sum of all four cells of a study’s fourfold table instead of only on the sum of events and non-events in a single treatment arm. In addition, we analysed in which situations ignoring randomization may be problematic.

The beta-binomial model respecting randomization performed well in our simulation study on meta-analyses of (very) rare events. The BBST showed effect estimates closer to the null, slightly lower bias on average, and converged more often compared to both common-beta models. Actually, we found an excess number of extreme values for bias. However, the respective estimates were almost always obviously implausible given the data of the individual studies included in the meta-analysis (e.g., each individual study showed a small effect but the pooled effect was very large) and the respective models would have been classified as not converged in practice anyway. In addition, this mainly happens when the effect sizes were very large (e.g., logOR > 5). In practice, in such a situation the choice of the synthesis method unlikely has an impact on the conclusion of the existence of an effect or would be an indication for a spurious result that would normally lead to the decision to perform sensitivity analysis using another meta-analysis method.

We calculated the MHKPM as a reference, particularly to check if the results are closer to the BBCBrespect model. However, we could not observe such a tendency. The reason could be that the similarity regarding respecting randomization is superimposed by other differences, particularly disregarding double zero studies and the continuity correction applied to single zero studies. As the odds ratios from the MHKPM model suffered much more from bias than those from the beta-binomial models and the very low power, the MHKPM model appears not to be a good alternative for (very) few events, anyway.

The BBCBrespect model respects randomization by conditioning on the total counts in each study. Noticeably, the individual study effects are not fully removed because all studies share the same beta, which means that information between trials is utilized. The comparison of the data situations where BBCBignor and BBCBrespect provide very different results revealed that this was mostly in cases when $\tau^2$was high, indicating that ignoring randomization is probably particularly problematic when the heterogeneity between the included studies is high. In some cases, the effects even changed sign. The observed differences might be due to the Simpsons-Paradox caused by missing stratification according to the individual study level characteristics [22]. For application in practice, this means that using the BB models ignoring randomization should be carefully considered when heterogeneity or study-level confounding (i.e., ecological bias) can be expected. However, as the risk of the Simpsons-Paradox is mainly relevant when studies with unbalanced samples are involved, caution is especially needed when studies with unbalanced randomizations (e.g., 2:1) are included. In addition, as can be seen in the log-likelihood function of BBCBignor, the estimation is not independent from the number of observations in one study arm. Therefore, BBST and BBCBignor should not be used when studies with unbalanced allocation ratios are included in a meta-analysis.

Similar to arm-based methods using random intercepts for the studies compromises the principle of concurrent control because the mean is shrunken to a common effect. Theoretically, it is also possible to implement a categorical fixed effect for the studies in the BB-models. Implementing a fixed-effect for the study intercept would fully respect the principle of concurrent control [23]. However, we and others have observed that using a fixed effect for the baseline risk may cause convergence problems which are likely because modelling a categorical fixed study effect implies estimating (K-1) parameters for modelling the study effect [24]. In addition, possible alternatives (inverse-variance fixed-effect models, Bayesian analysis, narrative synthesizes) also come along with limitations (assuming heterogeneity, assumptions about priors, subjectivity). An advantage of the standard and the proposed BB models is that generally only three parameters must be estimated. Particularly in very sparse data situations, it is reasonable to make the best possible of all available information thus utilizing between study information, for example by modelling a random effect, appears reasonable. Likewise, our Cochrane scenario indicates that in the case of few studies and rare events, arm-based synthesis (ignoring randomization), compared to contrast-based synthesis (respecting randomization) usually results in similar point estimates but leads to a slight increase in power. Nevertheless, it is not well explored yet, in which situations borrowing information between studies may bias the results and thus some caution is needed.

The idea of using panel data estimation routines for estimating BBCBs and respecting randomization by conditioning on all four cells of the study's fourfold table for estimating summary estimates opens the door for using other panel count data models as commonly used in econometrics such as the FE-Poisson, RE-Poisson, or RE-Negbin models [15]. The appeal of these models is that they all have a closed-form likelihood and can be conveniently implemented using standard procedures from statistical software as used in econometrics (e.g. PROC COUNTREG SAS, plm package in R, xtnbreg in Stata).

**Conclusion**

The results show that using the standard BBM, which ignores randomization is usually not biased when the randomization is balanced and the size of studies included in the meta-analysis is not very different. However, as possible ecological bias due to ignoring randomization is an inherent disadvantage of the model and the BBM that respects randomization shows very similar results in the simulation study, the latter may be generally preferred.

## Tables

Table 1: Log-likelihood functions for the BBCBignor and the BBCBrespect model. The difference between both likelihood functions is in the first six terms of the BBCBignor log-likelihood as compared to the first three terms in the BBCBrespect loglike-lihood, whereas the last 12 terms in both log-likelihoods are identical and therefore set in grey color. “lgamma” denotes the logarithm of the gamma function.

| BBCBignor | BBCBrespect |
|---|---|
| $ll(\alpha_C, \alpha_T, \beta) =$ | $ll\ (\alpha_C, \alpha_T, \beta) =$ |
| $lgamma(n_C+1) +$<br>$lgamma(n_T+1) +$ | $lgamma(n_C + n_T + 1) +$ |
| $lgamma(\alpha_C+ \beta) +$<br>$lgamma(\alpha_T+ \beta) -$ | $lgamma(\alpha_C + \beta + \alpha_T + \beta) -$ |
| $lgamma(n_C+\alpha_C+ \beta) -$<br>$lgamma(n_T+\alpha_T+ \beta ) -$ | $lgamma(\alpha_C + \beta + \alpha_T + \beta + n_C + n_T) -$ |
| $lgamma(y_C+1) - lgamma(n_C - y_C+1) - lgamma(y_T+1) - lgamma(n_T - y_T+1) +$ | $lgamma(y_C+1) - lgamma(n_C - y_C+1) - lgamma(y_T+1) - lgamma(n_T - y_T+1) +$ |
| $lgamma(y_C+\alpha_C) + lgamma(n_C - y_C+\beta) + lgamma(y_T+\alpha_T) + lgamma(n_T - y_T+ \beta) -$ | $lgamma(y_C+\alpha_C) + lgamma(n_C - y_C+\beta) + lgamma(y_T+\alpha_T) + lgamma(n_T - y_T+ \beta) -$ |
| $lgamma(\alpha_C) - lgamma(\beta) - lgamma(\alpha_T) - lgamma(\beta)$ | $lgamma(\alpha_C) - lgamma(\beta) - lgamma(\alpha_T) - lgamma(\beta)$ |

Table 2: Description of the simulation study

| **Parameter** | **Distributional assumption and parameter specification** |
|---|---|
| Sample size of single study $n_k$ | Generated from a log-normal distribution with $\mu_{ND}$ = 4.615 and $\sigma_{ND}$ = 1.1 |
| Sample size of treatment (control) arm of single study $n_{kT}$ ($n_{kC}$) | For $n_{kT}$: Generated from a binomial distribution with event probability 0.5 (1:1 randomization) and $n_k$ as number of experiments<br>For $n_{kC}$: $n_k - n_{kT}$<br>Studies we a sample size <10 in one of the arms were excluded |
| Variation $\sigma^2$ within study | Is implicitly given by the random sample size of a single study and event probability in the control group |
| Effect size of $\theta$ = logOR under $H_1$ for the pooled common effect | Generated from a log normal distribution with $\mu_{ND}$ = –0.59, $\sigma_{ND}$ = 0.61, skewness = –1.28 and kurtosis = 3.68<br>Fleishman power transformation (23) was used to for generating the skwed distribution |
| Heterogeneity $\tau^2$ of the common effect between the studies (for $\theta$ = log OR) | Generated from a log normal distribution with $\mu_{ND}$ = –1.47, $\sigma_{ND}$ = 1.65 and skewness = –0.55 using Fleishman's power transformation to generate the skewed distribution (23) |
| Event probability in treatment group $\pi_{T,true}$ | Generated from the effect size |
| Events in the intervention and control group | Generated from a binomial distribution with $\pi_{C,true}$ ($\pi_{T,true}$), and $n_{kT}$ ($n_{kC}$) |

Table 3: Proportion of zero studies for each scenario

| **Scenario** | | | **Proportion of zero studies** | | | |
|---|---|---|---|---|---|---|
| ***Number of studies in meta-analysis*** | ***Effect*** | ***Baseline probability in the control arm*** | ***Proportion of zero studies in treatment arm*** | ***Proportion of zero studies in control arm*** | ***Proportion of single zero studies in either control or treatment arm*** | ***Proportion of double zero studies*** |
| "CO" | "H0" | 0.01 | 0.54 | 0.55 | 0.73 | 0.36 |
| "CO" | "H0" | 0.02 | 0.38 | 0.37 | 0.55 | 0.2 |
| "CO" | "H0" | 0.05 | 0.19 | 0.17 | 0.29 | 0.06 |
| "CO" | "H1" | 0.01 | 0.62 | 0.55 | 0.77 | 0.4 |
| "CO" | "H1" | 0.02 | 0.47 | 0.37 | 0.61 | 0.24 |
| "CO" | "H1" | 0.05 | 0.26 | 0.16 | 0.35 | 0.08 |
| "NC" | "H0" | 0.01 | 0.54 | 0.55 | 0.73 | 0.36 |
| "NC" | "H0" | 0.02 | 0.38 | 0.37 | 0.55 | 0.2 |
| "NC" | "H0" | 0.05 | 0.19 | 0.16 | 0.29 | 0.06 |
| "NC" | "H1" | 0.01 | 0.62 | 0.55 | 0.77 | 0.4 |
| "NC" | "H1" | 0.02 | 0.47 | 0.37 | 0.6 | 0.24 |
| "NC" | "H1" | 0.05 | 0.27 | 0.16 | 0.35 | 0.08 |

Table 4: Two simulated meta-analyses for illustrating meta-analyses with differing results between BBCBignor and BBCBrespect

| | | | | | | **Meta-analysis results** | | | |
|---|---|---|---|---|---|---|---|---|---|
| | **Study ID** | **$n_T$** | **$n_C$** | **$y_T$** | **$y_C$** | $\tau^2$ | **LogOR BBCBignor** | **LogOR BBCBrespect** | **LogOR MKHPM** |
| **Example different study sizes** | | | | | | | | | |
| | 1 | 11 | 10 | 0 | 0 | 0.00001 | 4,04312E+14 | -0.2877 | -0.2655 |
| | 2 | 23 | 16 | 1 | 2 | | | | |
| | 3 | 10 | 14 | 0 | 0 | | | | |
| | 4 | 55 | 55 | 0 | 0 | | | | |
| | 5 | 293 | 321 | 5 | 5 | | | | |
| | 6 | 12 | 11 | 0 | 0 | | | | |
| | 7 | 36 | 34 | 0 | 1 | | | | |
| **Example heterogenity** | | | | | | | | | |
| | 1 | 255 | 255 | 9 | 3 | 1,51037E+14 | -0.7100 | 7,45498E+13 | -0.3160 |
| | 2 | 166 | 160 | 0 | 3 | | | | |
| | 3 | 98 | 86 | 0 | 1 | | | | |

## Figures

Figure 1: Different summarization levels of the BB models and the consequences of respecting or ignoring randomization

| Study | Treatment/ Control | Study Group | Success (events) | Count Success |
|---|---|---|---|---|
| Study 1 | T | 1 | Yes | 2 |
| Study 1 | T | 1 | No | 140 |
| Study 1 | C | 2 | Yes | 5 |
| Study 1 | C | 2 | No | 134 |
| Study 2 | T | 3 | Yes | 3 |
| Study 2 | T | 3 | No | 197 |
| Study 2 | C | 4 | Yes | 6 |
| Study 2 | C | 4 | No | 195 |
| Study 3 | T | 5 | Yes | 2 |
| Study 3 | T | 5 | No | 98 |
| Study 3 | C | 6 | Yes | 2 |
| Study 3 | C | 6 | No | |
| … | … | … | … | … |

BBCB ignor

BBCB respect

BBST

Beta ($\alpha_T$, $\beta_T$)

Beta ($\alpha_C$, $\beta_C$)

Brackets indicate the unit on which the maximum likelihood function is conditioned; the sum sign Σ indicates the rows over which it is summed up; percent signs indicate the estimation of the proportion of events

Figure 2: converged runs under H1

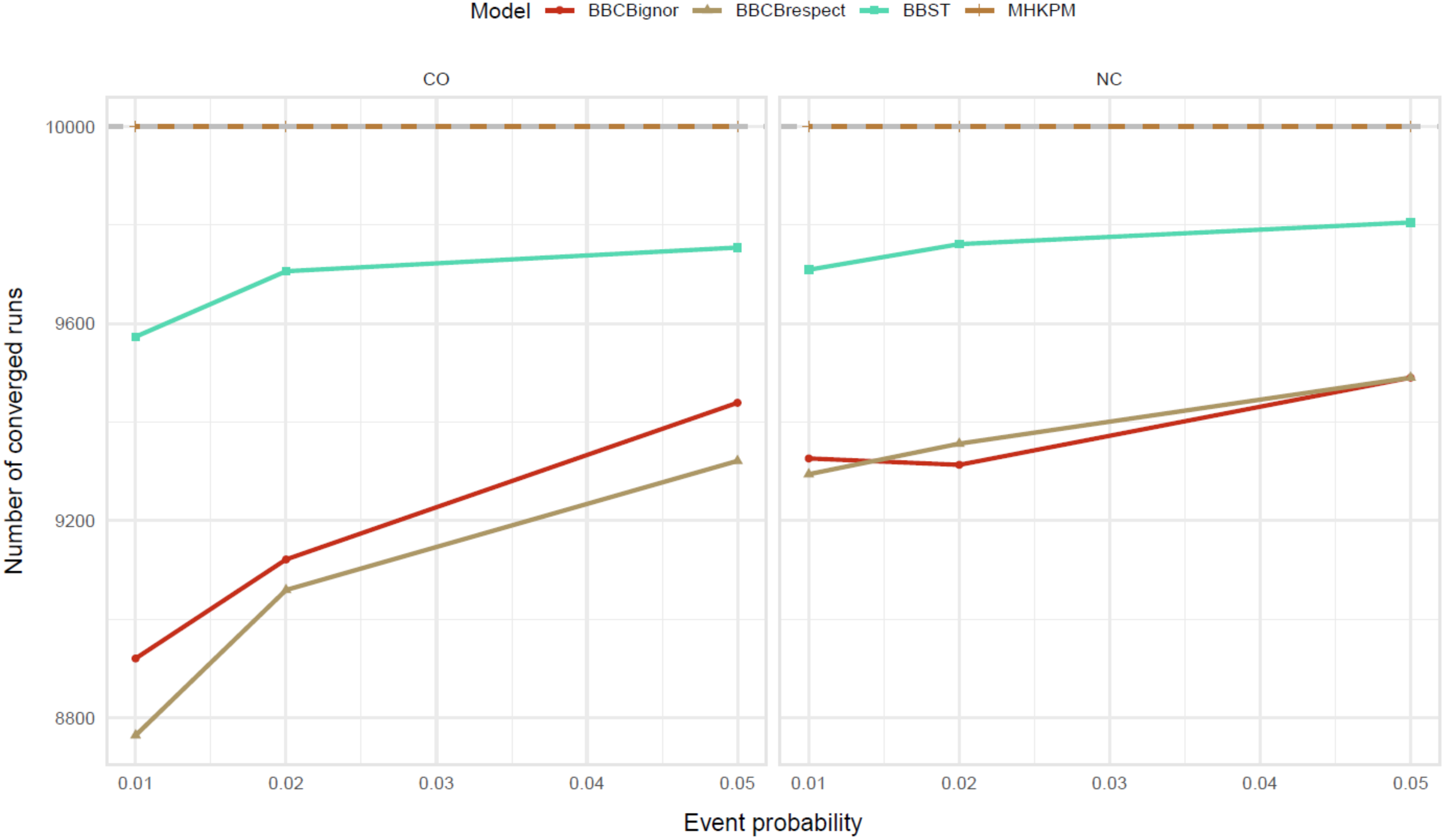

Figure 3: Bias under H1

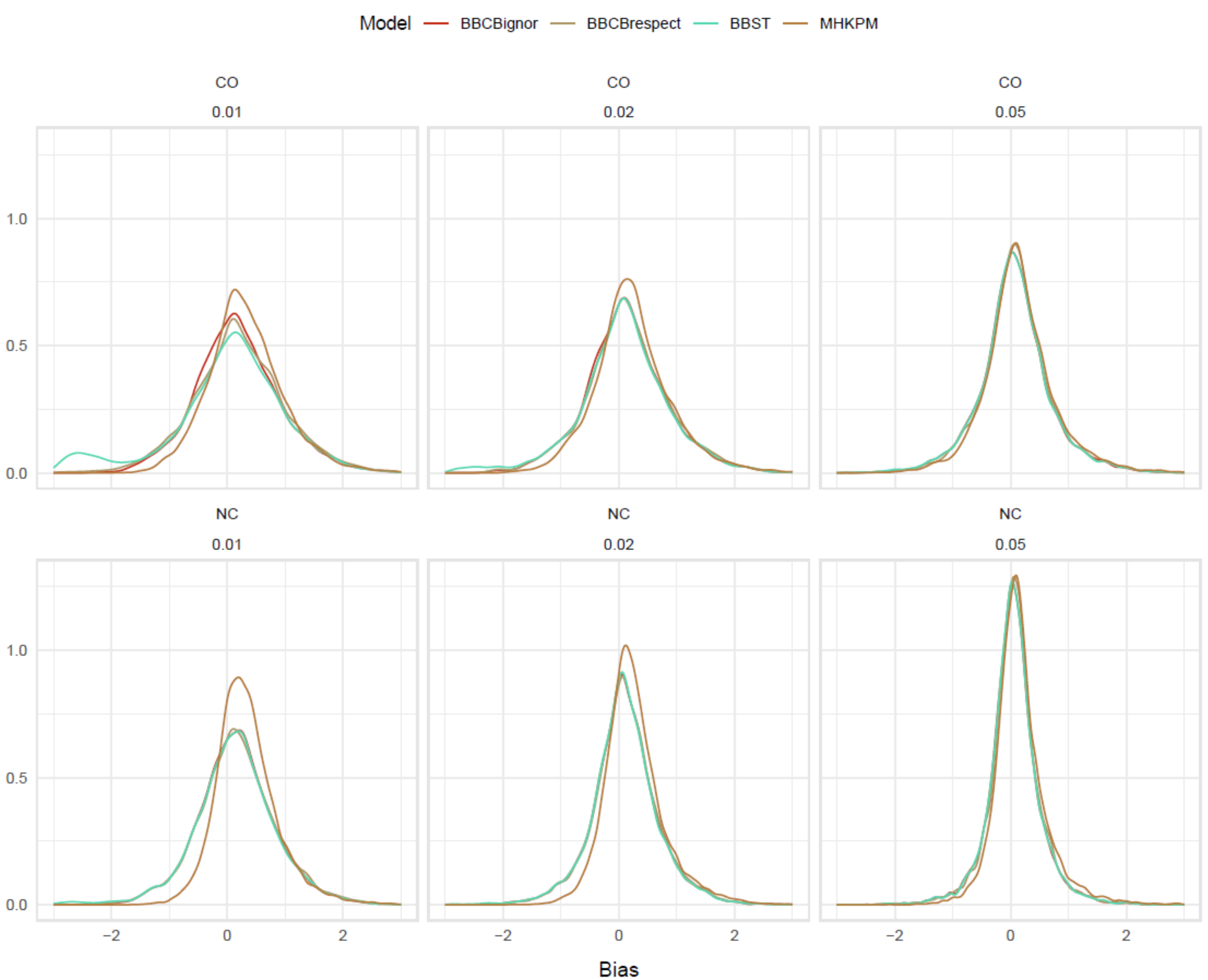

Figure 4: Coverage probability under H1

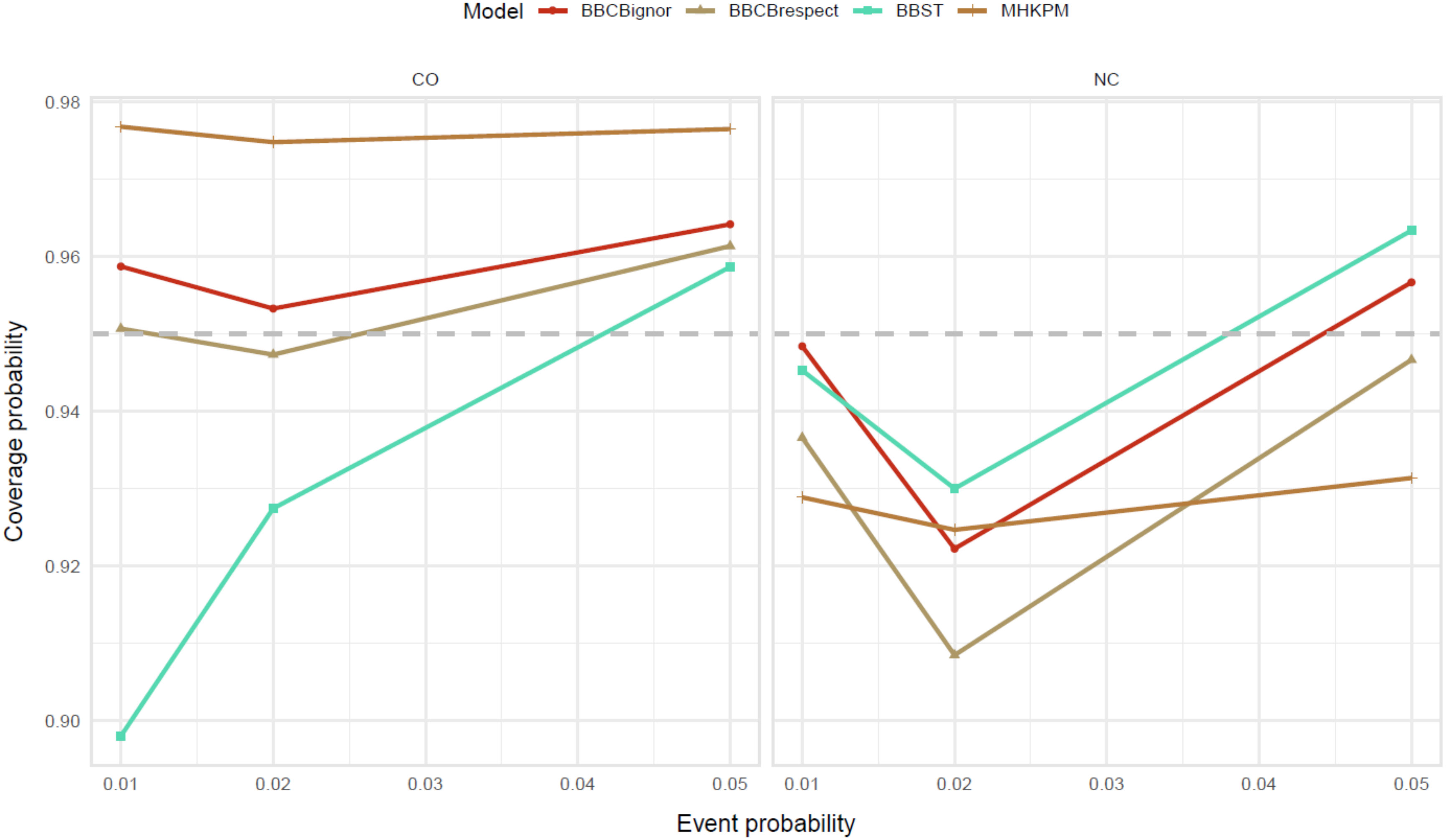

Figure 5: length of the 95%-confidence intervals under H1

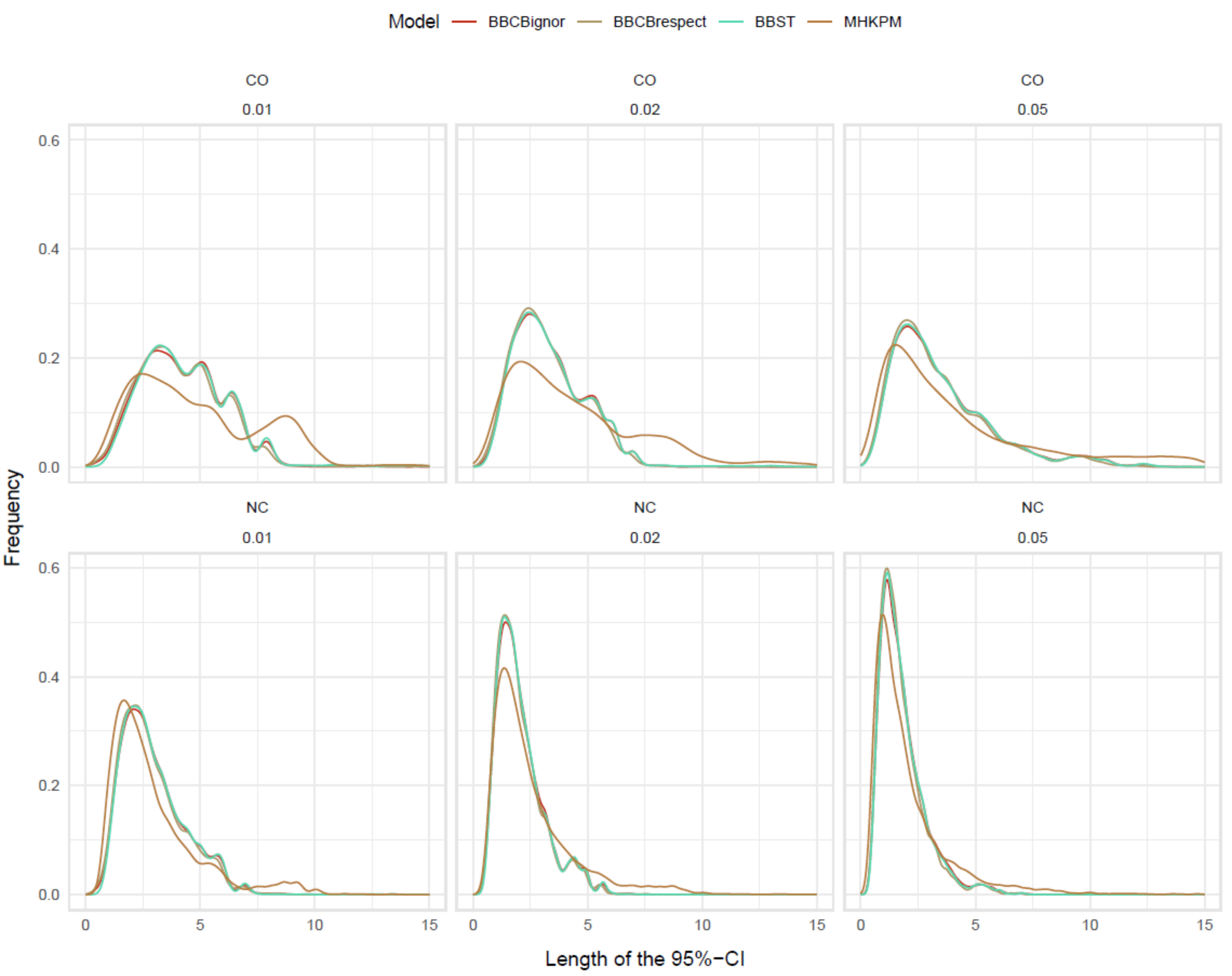

Figure 6: Power

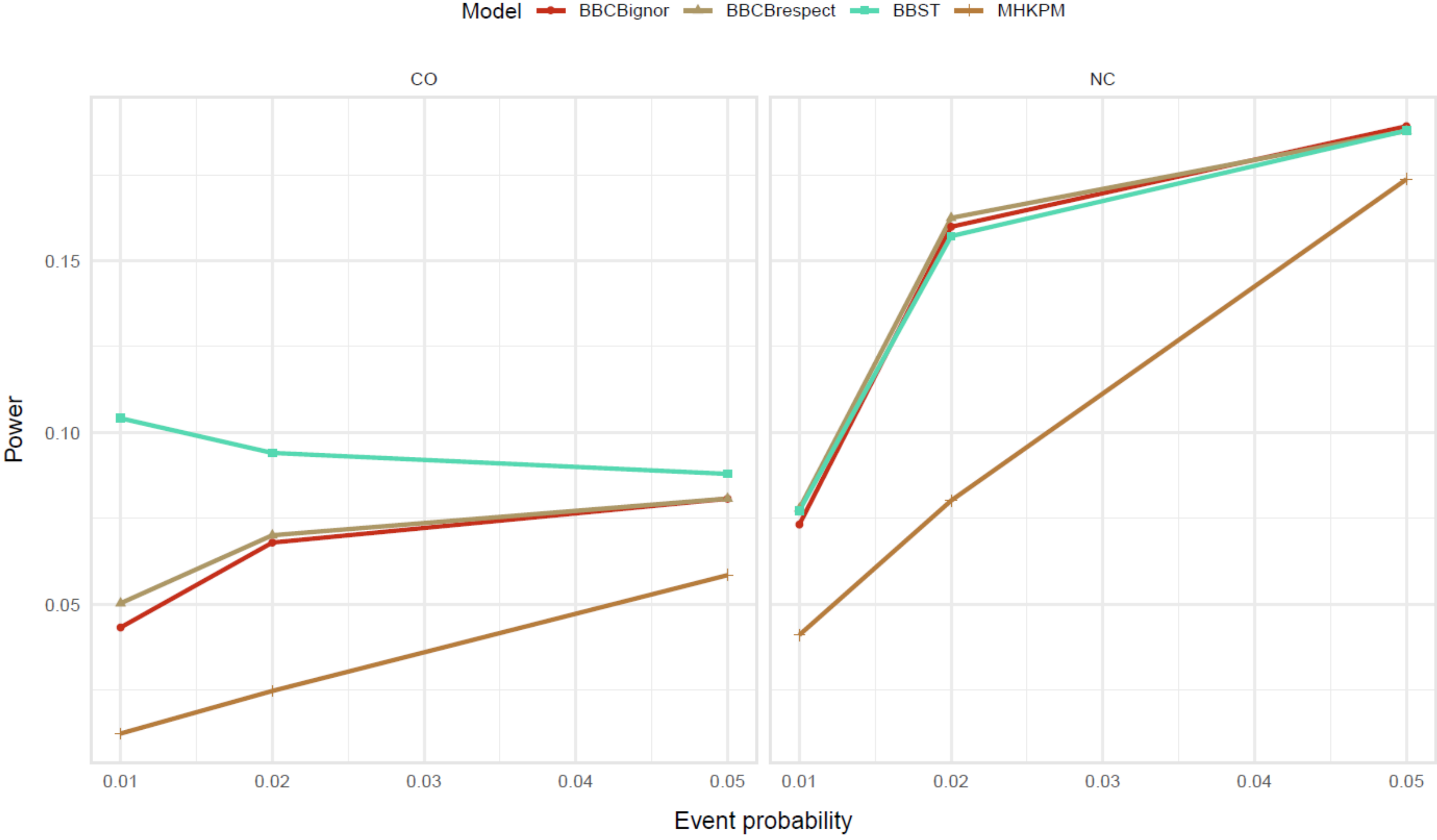

Figure 7: Bland-Altman-plot for the agreement of the logOR of BBST vs BBCBignor

CO

Difference of logOdds ratio

20

0

−20

−10 −5 0 5 10

Size of logOdds ratio

NC

20

0

−20

−10 −5 0 5 10

Size of logOdds ratio

The Bland-Altman plot shows the agreement (difference of logOdds ratios) of BBST compared to BBCBignor; figures were clipped-of at a logOR of 10

Figure 8: Bland-Altman-plot for the agreement of the logOR of BBCBignor vs. BBCBrespect

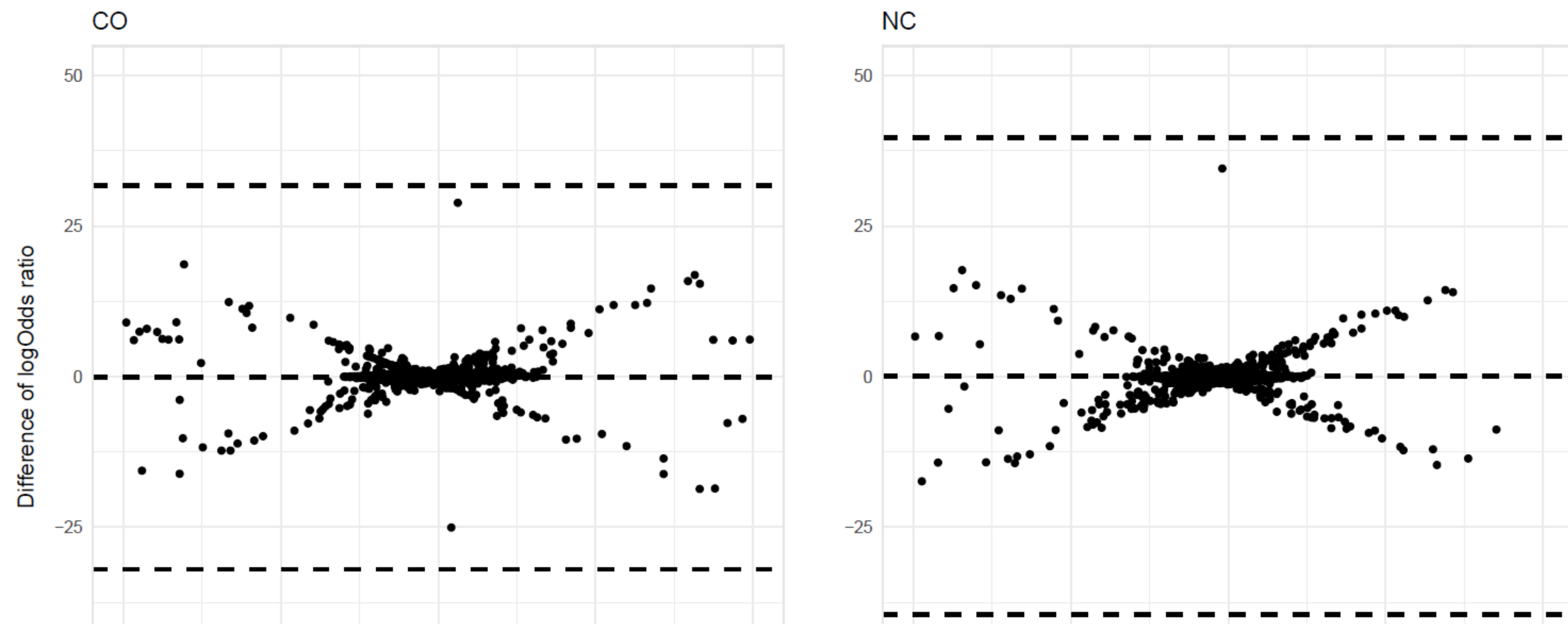

The Bland-Altman plot shows the agreement (difference of logOdds ratios) of BBCBignor compared to BBCBrespect; figures were clipped-of at a logOR of 10

Figure 9: example meta-analysis on on-pump versus off-pump artery bypass grafting

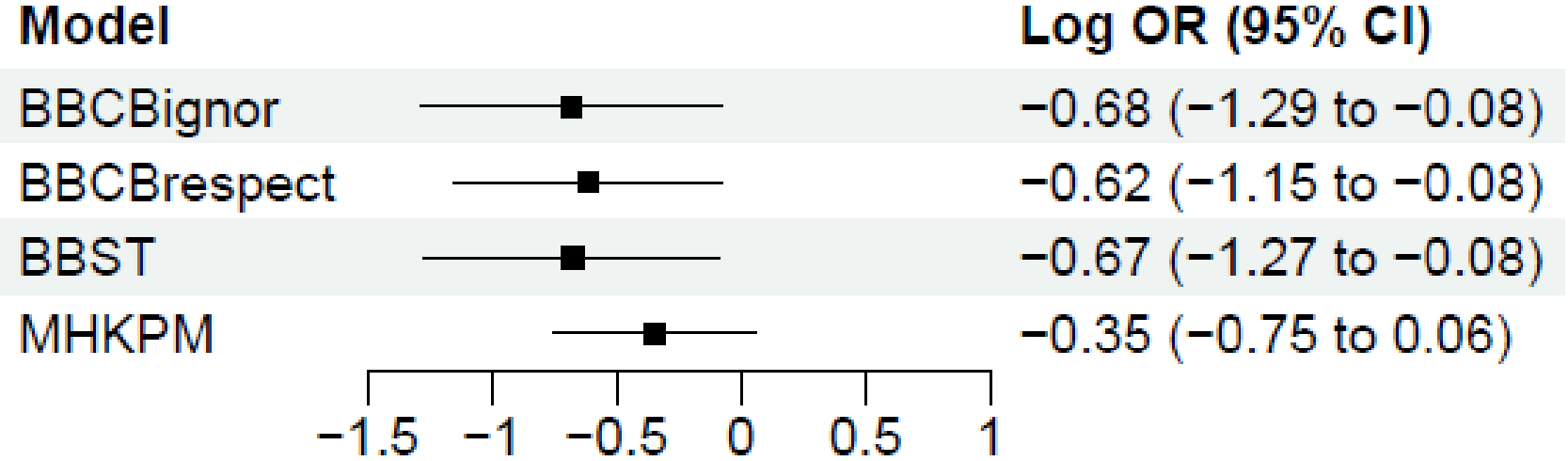

## Supplement

### Simulation-Code

**This program performs a simulation study for assessing the performance of common beta-binomial**

**in comparison to generalized linear mixed models and two-stage meta-analyis models.**

**It introduces the SAS macro %RegressionModelsSim which performs the simulation (data generation, computation of starting values, parameter estimation, and summarization**

**of results in a single data set) for the simulation scenarios. A single simulation scenario is defined by the macro parameters**

**effect : =1 H1 generate ORs**

**: =0 H0 no effect OR=1**

**rem : =1 Generate data from the standard random effects model**

**=0 Generate data from the standard fixed effects model**

**nsimruns : Number of generated meta-analyses**

**nstudy : Number of studies in a single meta-analysis**

**= 0 Generate this number from the distribution as given in Turner et al.,**

**= 1 Generate this number from the distribution as given in Page et al.,**

**= number Generate according the respective number**

**baselineprob : Baseline probability in the control group**

**= 0 Generate the baseline probaility according in Turner et al.,**

```
                                  = number Generate according the respective number
  seedsim        : Initializing value of the random number generator
  printsimdata     : =1  Print the simulated data
                =0  (Default) Do not print the simulated data



options pagesize=60 linesize=120;
libname result "C:\Users\mathes1\Desktop\SAS";
libname user "C:\Users\mathes1\Desktop\SAS";



%macro RegressionModelsSim(nstudy=,baselineprob=,effect=,rem=,seedsim=,nsimruns=,printsimdata=0);


***********************************************************************************;
*** PRELIMINARIES ********************************************************************;
***********************************************************************************;
```

```
options nomlogic nomprint nosource nosource2 nonotes nosymbolgen;
proc printto log="C:\Users\mathes1\Desktop\SAS\Log_&seedsim..txt"
          print="C:\Users\mathes1\Desktop\SAS\Output_&seedsim..txt";
run;



proc format;
    value treatf  1=" Treatment" 0="Control";
    value eventf  1=" Yes" 0="No";
    value  $ident_t 'H0'='No treatment effect' 'H1'='Average treatment effect';
run;


*********************************************************************************************************************************
*******************************************;
*********************************************************************************************************************************
*******************************************;
********** DATA GENERATION
*********************************************************************************************************************************
****************;
*********************************************************************************************************************************
*******************************************;
```

```
***************************************************************************************************************************************************************
****************************************************;

 * Define a id (IDENT) for the simulation scenarios;
 data ident;
    %if &effect = 0     %then %do;ident_t='H0';%end;
    %if &effect = 1           %then %do;ident_t='H1';%end;

    %if &rem = 0  %then %do;ident_REM='FEM';%end;
    %if &rem = 1  %then %do;ident_REM='REM';%end;




           %if &baselineprob = 0 %then %do;ident_BASELINEPROB='stand';%end;
           %if &baselineprob = 0.01 %then %do;ident_BASELINEPROB='01';%end;
           %if &baselineprob = 0.02 %then %do;ident_BASELINEPROB='02';%end;
           %if &baselineprob = 0.05 %then %do;ident_BASELINEPROB='05';%end;

           %if &nstudy=0  %then %do; ident_nstudy='CO';%end;
           %if &nstudy=1  %then %do; ident_nstudy='NC';%end;
```

```
    Ident =  ident_t || ident_BASELINEPROB || ident_REM || ident_nstudy;
    call symput("Ident",Ident);
 run;
 %put  Ident: &Ident;

* The data set SIMDATA contains the simulated data sets, the variable SIMRUNS indexes the single simulation run;
 data simdata;

   * Initialize random seed;
    call streaminit(&seedsim);



   ************************** Loop across the number of simulation runs ***************************************************************;
    do simruns=1 to &nsimruns;

    ***** Begin data generation ******************************************************************************************************;
```

*** ... Declare number of studies from the macro variable &nstudy;**

**NumberofStudies=&nstudy;**

*** Generate a baseline event probability from a beta distribution with alpha=0.393628 und beta=0.881246.**

**These values originate from an analysis of the 14,886 meta-analyses with binary outcomes from Rebecca Turner**

**(vgl. Analyse_Baseline_Risks_Turner.SAS)**

**A decent fit results from fitting a beta distribution with alpha=0.422932 and beta=1.433449. To be concrete,**

**comparing observed (from Turner) vs. fitted summary statistics, we find for the**

**Median 0.1262 vs. 0.1325, for the**

**Mean 0.2271 vs. 0.2336 and for the**

**standard deviation 0.2558 vs. 0.2556**

**The true mean and standard deviation from a Beta(0.422932, 1.433449) distribution are 0.22783 and 0.24817**

**ATTENTION: The fact that the baseline probabilities are simulated from a beta distribution DOES NOT MEAN that the**

**data are generated from the beta-binomial model. We here generate one fixed baseline probability for a single meta-analysis,**

**whereas the beta-binomial model assumes that the baseline probabilities OF THE STUDIES WITHIN THE META-ANALYSIS follow**

**a beta distribution.**

**The true model is still a standard inverse-variance random-effects model;**

```
*for the standard secenario the baseline probablity pc is sample from this beta destribution;
%if &baselineprob=0  %then %do;
true_pc=rand('BETA',0.422932,1.433449);
if true_pc<0.01 then do; true_pc=0.01; end;
true_logit_pc=log(true_pc/(1-true_pc));
%end;

%if &baselineprob=0.01  %then %do;
true_pc=0.01;
true_logit_pc=log(true_pc/(1-true_pc));
%end;

%if &baselineprob=0.02  %then %do;
true_pc=0.02;
true_logit_pc=log(true_pc/(1-true_pc));
%end;
```

```
                %if &baselineprob=0.05  %then %do;
                true_pc=0.05;
                true_logit_pc=log(true_pc/(1-true_pc));
%end;


              * For the H0 situation, the true OR is set to 1;
%if &effect=0 %then %do;
                true_OR=1;
                true_LogOR=0;
%end;


              * For the H1 situation, the ORs have to be generated;
%if &effect=1 %then %do;

  * Generation of the effects (ORs);
                           LogNormal_mean_Turner_OR= -0.59;LogNormal_stddev_Turner_OR=0.61;
  * Generation of a skewed distribution using Fleishmans power transformation method;
   random_standard_normal_OR=rand("NORMAL");
                           random_skewed_normal_OR  =  0.659546319              +  0.659546319*random_standard_normal_OR
```

```
                    -0.228926754*random_standard_normal_OR**2 -  0.087359908*random_standard_normal_OR**3;


 true_LogOR=LogNormal_mean_Turner_OR + LogNormal_stddev_Turner_OR*random_skewed_normal_OR;
          if true_OR<0.99 then do;
                true_OR=exp(true_LogOR);end;



          if true_OR>0.99 then do;
                LogNormal_mean_Turner_OR= -0.59;LogNormal_stddev_Turner_OR=0.61;


* Generation of a skewed distribution using Fleishmans power transformation method;
 random_standard_normal_OR=rand("NORMAL");
                random_skewed_normal_OR  =  0.659546319          +  0.659546319*random_standard_normal_OR
             -0.228926754*random_standard_normal_OR**2 -  0.087359908*random_standard_normal_OR**3;


 true_LogOR=LogNormal_mean_Turner_OR + LogNormal_stddev_Turner_OR*random_skewed_normal_OR;


                true_OR=exp(true_LogOR); end;


                              if true_OR>0.99 then do;
                LogNormal_mean_Turner_OR= -0.59;LogNormal_stddev_Turner_OR=0.61;
```

```
* Generation of a skewed distribution using Fleishmans power transformation method;
  random_standard_normal_OR=rand("NORMAL");
                         random_skewed_normal_OR  =  0.659546319               +  0.659546319*random_standard_normal_OR
                    -0.228926754*random_standard_normal_OR**2 -  0.087359908*random_standard_normal_OR**3;

  true_LogOR=LogNormal_mean_Turner_OR + LogNormal_stddev_Turner_OR*random_skewed_normal_OR;

                         true_OR=exp(true_LogOR); end;

                if true_OR>0.99 then do;
                         LogNormal_mean_Turner_OR= -0.59;LogNormal_stddev_Turner_OR=0.61;

* Generation of a skewed distribution using Fleishmans power transformation method;
  random_standard_normal_OR=rand("NORMAL");
                         random_skewed_normal_OR  =  0.659546319               +  0.659546319*random_standard_normal_OR
                    -0.228926754*random_standard_normal_OR**2 -  0.087359908*random_standard_normal_OR**3;

  true_LogOR=LogNormal_mean_Turner_OR + LogNormal_stddev_Turner_OR*random_skewed_normal_OR;

                         true_OR=exp(true_LogOR); end;
```

```
if true_OR>0.99 then do;
        LogNormal_mean_Turner_OR= -0.59;LogNormal_stddev_Turner_OR=0.61;

* Generation of a skewed distribution using Fleishmans power transformation method;
 random_standard_normal_OR=rand("NORMAL");
        random_skewed_normal_OR  =  0.659546319          +  0.659546319*random_standard_normal_OR
      -0.228926754*random_standard_normal_OR**2 -  0.087359908*random_standard_normal_OR**3;

 true_LogOR=LogNormal_mean_Turner_OR + LogNormal_stddev_Turner_OR*random_skewed_normal_OR;

        true_OR=exp(true_LogOR); end;

if true_OR>0.99 then do;
    true_LogOR=log(0.684); true_OR=0.684;
    end;

%end;
```

```
* For the random effects situation, the tau-squares have to be generated;
 %if &rem=1 %then %do;

   * Generation of the random effects variance (tau-square);
    LogNormal_mean_Turner_Tau=-1.47;LogNormal_stddev_Turner_Tau=1.65;

   * Generation of a skewed distribution using Fleishmans power transformation method;
    random_standard_normal_tau=rand("NORMAL");
    random_skewed_normal_tau  =  0.104796973                        +  1.049992872*random_standard_normal_tau
                    -0.104796973*random_standard_normal_tau**2 -  0.020782441*random_standard_normal_tau**3;

    Log_TauSquare_skewed=LogNormal_mean_Turner_Tau + LogNormal_stddev_Turner_Tau*random_skewed_normal_tau;
    true_TauSquare=exp(Log_TauSquare_skewed);
 %end;

            * The number of studies is generated from a "Cochrane" scheme which mirrors the distribution in Turner et al. and a
             "Non-Cochrane" scheme which mirrors the distribution in Page et al.
             &nstudy=0  reproduces the Turner distribution,
             &nstudy=1 reproduces the Page distribution,
             &nstudy given as a positive integer generates meta-analysis with a fixed number of studies;
             %if &nstudy=0 %then %do;
```

```
                LogNormal_mean_Turner_NStud=0.65; LogNormal_stddev_Turner_NStud=1.2;
        Log_NumberofStudies_Turner = LogNormal_mean_Turner_NStud + LogNormal_stddev_Turner_NStud*rand('normal');
    NumberofStudies=1+ceil(exp(Log_NumberofStudies_Turner));
  * To avoid simulation time being too log, the number of studies is limited to 50;
                     if NumberofStudies>=50 then NumberofStudies=50;
                     if NumberofStudies<2 then NumberofStudies=2;
                     call symput('nstudy',NumberofStudies);
              %end;

              %if &nstudy=1 %then %do;
                LogNormal_mean_Page_NStud=1.94; LogNormal_stddev_Page_NStud=0.925;
        Log_NumberofStudies_Page = LogNormal_mean_Page_NStud + LogNormal_stddev_Page_NStud*rand('normal');
    NumberofStudies=1+ceil(exp(Log_NumberofStudies_Page));
                    * To avoid simulation time being too log, the number of studies is limited to 50;
                     if NumberofStudies>=50 then NumberofStudies=50;
                     if NumberofStudies<2 then NumberofStudies=2;
                     call symput('nstudy',NumberofStudies);
              %end;


******************* Loop across the number of studies within a meta-analysis *************************************************;
```

```
do study=1 to NumberofStudies;
 * Generate the sample size for the two groups following the distribution in Turner et al., Table 2;
  LogNormal_mean_Turner_SSize=4.615;LogNormal_stddev_Turner_SSize=1.1;
  Log_StudySize=LogNormal_mean_Turner_SSize + LogNormal_stddev_Turner_SSize*rand('NORMAL');


  *Determine the study size by doubling the "ceiled" group size, where ceiling means
  calculating the smallest integer that is greater than or equal to the argument;

     StudySize=1+ceil(exp(Log_StudySize));
                                   if StudySize<20 then do; Studysize=20; end;


 * In the random effects situation, the true event probabilities are varying according to the random Log Odds Ratio;
  %if &rem=1 %then %do;
     true_TauSquare_i=true_TauSquare*rand('NORMAL');


    * Calculate the study-specific event probability in the treatment group on the expit scale.
     The linear predictor here is the sum of the logit of the event probability in the control group, the true LogOR und the random tau-square in the
     respective study;
```

```
    true_pt_i=exp(true_logit_pc + true_LogOR + true_TauSquare_i)/(1+exp(true_logit_pc + true_LogOR + true_TauSquare_i));
 %end;

 %if &rem=0 %then %do;
    true_pt=exp(true_logit_pc + true_LogOR)/(1+exp(true_logit_pc + true_LogOR));
 %end;

* Simulate the sample size in the treatment group by mimicking a randomization process with
 a binomial probability 0.5;
 nt=rand('BINOMIAL',0.5,StudySize);
 nc=StudySize-nt;

 * Exclude extrem cases with <10 treated observations in a treatment group;
 if nt<10 then do; nt=10; end;
 if nc<10 then do; nc=10; end;

* Generate the number of events in both groups;
                    sc=rand('BINOMIAL',true_pc,nc);

 %if &rem=0 %then %do;st=rand('BINOMIAL',true_pt,nt);    %end;
 %if &rem=1 %then %do;st=rand('BINOMIAL',true_pt_i,nt);  %end;
```

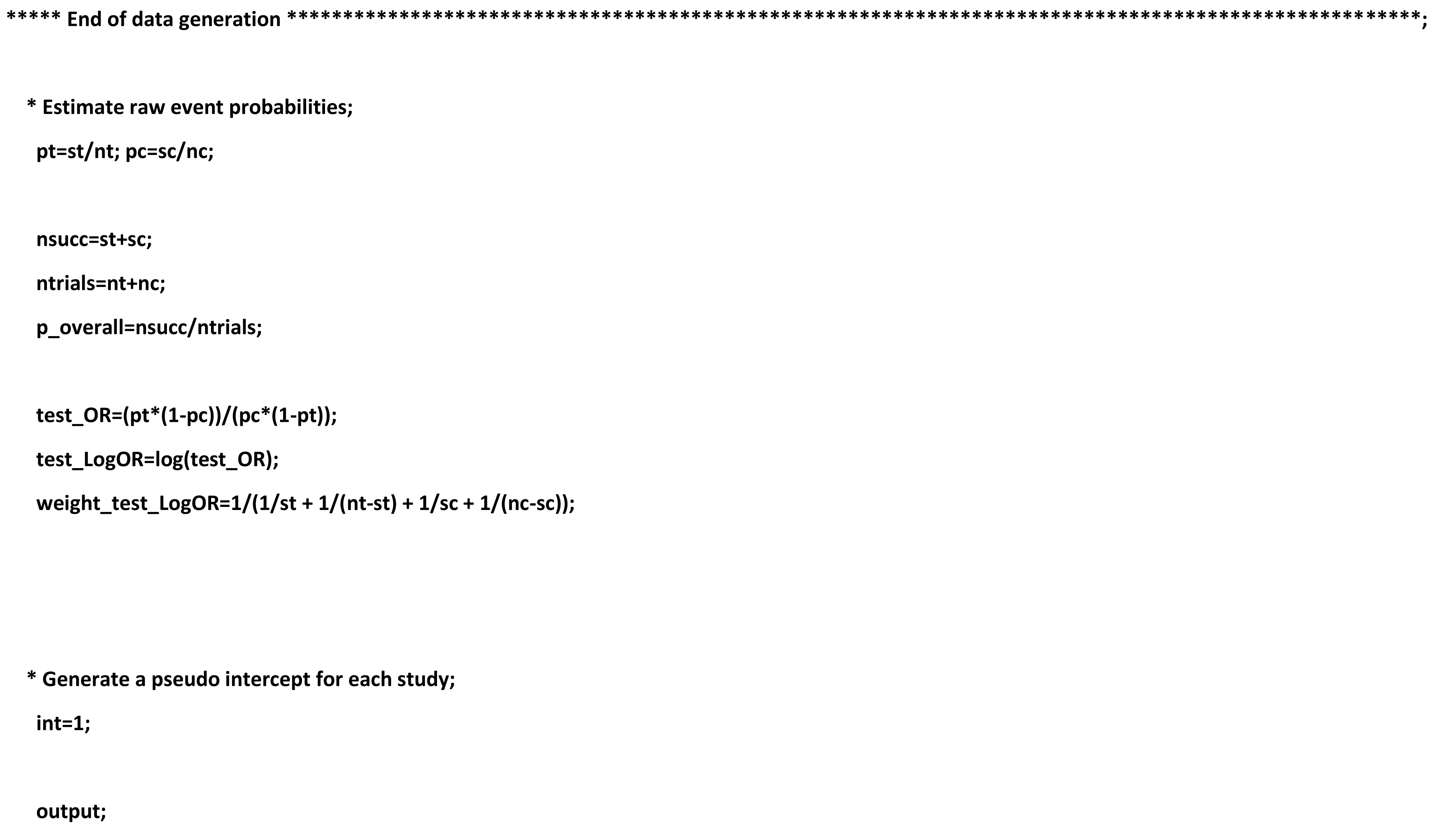

```
***** End of data generation ****************************************************************************************************************;

 * Estimate raw event probabilities;
  pt=st/nt; pc=sc/nc;

  nsucc=st+sc;
  ntrials=nt+nc;
  p_overall=nsucc/ntrials;

  test_OR=(pt*(1-pc))/(pc*(1-pt));
  test_LogOR=log(test_OR);
  weight_test_LogOR=1/(1/st + 1/(nt-st) + 1/sc + 1/(nc-sc));



 * Generate a pseudo intercept for each study;
  int=1;

  output;
```

```
        end; * of loop across studies within a meta-analysis;
      end; * of loop across the number of simulation runs;
  run;

  %if &printsimdata=1 %then %do;
     proc print data=simdata;
         title"simdata";
     run;
  %end;

* Control data generation;
* ods select none;
  /*proc means data=simdata n mean std median min q1 q3 max;
      %if &rem=1 %then %do; var nt nc StudySize st sc pt pc test_OR true_pc true_OR true_LogOR true_TauSquare_i  true_pt_i; %end;
      %if &rem=0 %then %do; var nt nc StudySize st sc pt pc test_OR true_pc true_OR true_LogOR; %end;
      title"Control generation of random effects data";
  run;*/
* ods select all;
```

```
* Reorganize SIMDATA, so that each single study has two observations (one for treatment, one for control, SIMDATADOUBLE) or
 each individual constitutes a single observation (SIMDATAEXPLODED);

 data simdatadescriptive;
             set simdata(keep=simruns true_LogOR);
             by simruns;
             if first.simruns;
 run;
 /*proc print data=simdatadescriptive;title"simdatadescritptive"; run;*/

 data simdatadouble_temp1;
     set simdata(keep=simruns study st sc nt nc);
             by simruns;
             do k=1 to 2;output;end;
 run;


 data simdatadouble(keep=simruns study treatment control success n logn dummy);
     set simdatadouble_temp1;
     by simruns;
     if mod(k,2)=1 then do; treatment=1; control=0; success=st; n=nt; end;
```

```
    if mod(k,2)=0 then do; treatment=0; control=1; success=sc; n=nc; end;
            dummy=1;
            logn=log(n);
run;

run;
/*proc print data=simdatadouble;title"simdatadouble";run;*/



* Data for the models that run in countreg;
data simdatahelp;
   set simdata(keep=simruns study st sc nt nc);
           by simruns;
                              do i=1 to 4; output; end;
                 dummy=1;
     run;



data simdataquadruple;
   set simdatahelp;
           by simruns;
```

```
    if i=1 then do; success=st;          treatment=1;indicator=1;CountregTreatment=treatment*(indicator=1);studygroup=study*2-1;end;
    if i=2 then do; success=nt-st;                      treatment=1;indicator=0;CountregTreatment=treatment*(indicator=1);studygroup=study*2-1;end;
          if i=3 then do; success=sc;          treatment=0;indicator=1;CountregTreatment=treatment*(indicator=1);studygroup=study*2;end;
    if i=4 then do; success=nc-sc;           treatment=0;indicator=0;CountregTreatment=treatment*(indicator=1);studygroup=study*2;end;
  run;
/*proc print data=simdataquadruple;title"simdataquadruple";run;*/


*Data for the inverse-variance models (with continuity correction);
  data simdatacontinuity(keep=simruns study st nt sc nc true_OR test_OR test_LogOR weight_test_LogOR int);
           set simdata;
           by simruns;
       if st=0 then do; st=0.5; sc=sc+0.5; end;
                   if sc=0 then do; sc=0.5; st=st+0.5; end;

                          pt=st/nt; pc=sc/nc;
        test_OR=(pt*(1-pc))/(pc*(1-pt));
        test_LogOR=log(test_OR);
        weight_test_LogOR=1/(1/st + 1/(nt-st) + 1/sc + 1/(nc-sc));
  run;
  /*proc print data=simdatacontinuity;title"simdatacontinuity";run;*/
```

```
*********************************************************************************************;
***  STARTING VALUES  **********************************************************************;
*********************************************************************************************;
* Compute means, standard deviations for the estimated event probabilities in both treatment groups;
 ods select none;
 proc corr data=simdata pearson cov;
     var pt;with pc;
             by simruns;
             ods output SimpleStats=SimpleStats_pt(keep=Simruns Variable Mean StdDev where=(Variable="pt") rename=(mean=start_pt
StdDev=start_stddev_pt))
                                    SimpleStats=SimpleStats_pc(keep=Simruns Variable Mean StdDev where=(Variable="pc") rename=(mean=start_pc
StdDev=start_stddev_pc));

 run;
 ods select all;
 /*proc print data=SimpleStats_pc;run;*/

 ods select none;
 proc means data=simdata mean stddev;
```

```
    var test_LogOR;
          by simruns;
    output out=SimpleStats_OR mean=start_LogOR StdDev=start_stddev_LogOR;
 run;
 ods select all;
/*proc print data=SimpleStats_OR;run;*/



* Starting values for the Beta-Binomial-Models;
 data startingvalues_Beta_Binomial(keep=simruns rho logalpha logbeta bsuccess b0 b_treat sigma);
    merge SimpleStats_pt(keep=simruns start_pt start_stddev_pt)
        SimpleStats_pc(keep=simruns start_pc start_stddev_pc)
                              SimpleStats_OR(keep=simruns start_LogOR start_stddev_LogOR);

   * Pooling variances and raw probabilities;
    if start_stddev_pt=0 then do; start_stddev_pt=0.001;end;
    if start_stddev_pc=0 then do; start_stddev_pc=0.001;end;
    if start_pt=0 then do; start_pt=0.001;end;
    if start_pc=0 then do; start_pc=0.001;end;

    pooledvar=(start_stddev_pt**2 + start_stddev_pc**2)/2;
```

```
    pooledp  =(start_pt+start_pc)/2;



    rho=pooledvar/(pooledp*(1-pooledp))/(1-pooledvar/(pooledp*(1-pooledp)));
        logitrho=log(rho/(1-rho));
    b0 =log(start_pc/(1-start_pc));
    if (start_pt=0) or (start_pc=0) then do; b_treat=0;                    end;
                else do; b_treat=log((start_pt*(1-start_pt))/(start_pc*(1-start_pc))); end;

        mu=exp((b0+b_treat)/(1+exp(b0+b_treat)));
    alpha=mu*((1-exp(logitrho))/(exp(logitrho)));
        logalpha=log(alpha);
    beta=(1-mu)*(1-exp(logitrho))/exp(logitrho);
        logbeta=log(beta);
        bsuccess=logalpha - logbeta;
        sigma=start_stddev_LogOR;

    by simruns;
  run;
/*proc print data=startingvalues_Beta_Binomial;title"startingvalues_Beta_Binomial";run;*/
```

```
************************************************************************************************************************************
*******************************************;
************************************************************************************************************************************
*******************************************;
********** ANALYSIS
************************************************************************************************************************************
**********************;
************************************************************************************************************************************
*******************************************;
************************************************************************************************************************************
*******************************************;
```

*** Akronyms for meta-analyses models:**

**BBST_ : Standard (common-rho) beta-binomial model(NLMIXED)**

**BBCB1 : Common-beta beta-binomial model disrespecting the randomization level(COUNTREG)**

**BBCB2 : Common-beta beta-binomial model respecting the randomization level (COUNTREG)**

**GLRR_ : Generalised linear mixed model with random intercept and random treatment effect (GLIMMIX)**

**HKPM_ : Knapp-Hartung-Method without variance correction (Paule-Mandel-heterogenity variance estimator)**

**MHKPM : Knapp-Hartung-Method with variance correction depedning on DL 95%-CIs (Paule-Mandel-heterogenity variance estimator)**

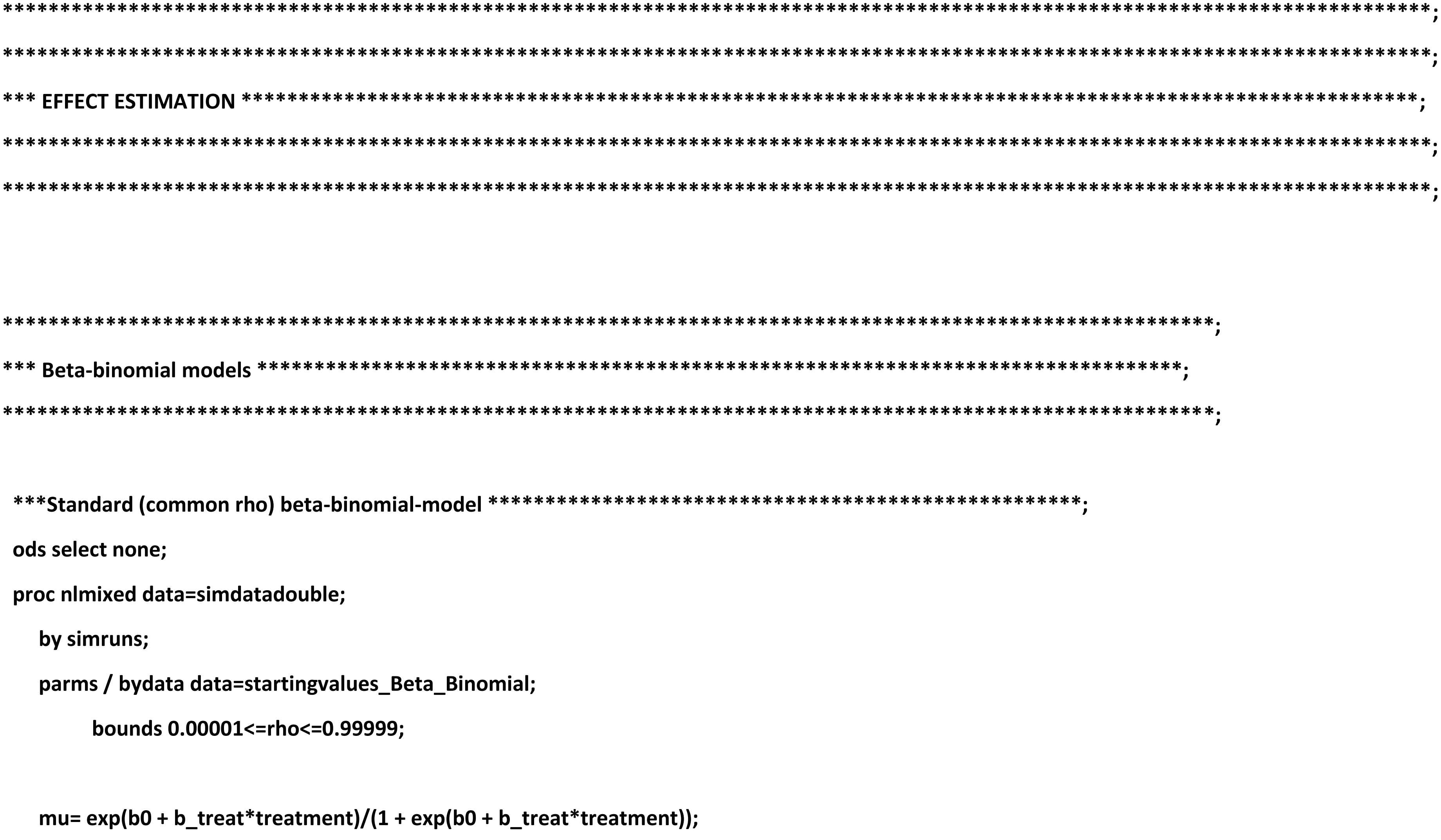

```
*************************************************************************************************************************************;
*************************************************************************************************************************************;
*** EFFECT ESTIMATION *************************************************************************************************************;
*************************************************************************************************************************************;
*************************************************************************************************************************************;

*****************************************************************************************************************;
*** Beta-binomial models **************************************************************************************;
*****************************************************************************************************************;

 ***Standard (common rho) beta-binomial-model *******************************************************;
 ods select none;
 proc nlmixed data=simdatadouble;
     by simruns;
     parms / bydata data=startingvalues_Beta_Binomial;
             bounds 0.00001<=rho<=0.99999;

     mu= exp(b0 + b_treat*treatment)/(1 + exp(b0 + b_treat*treatment));
```

```
    alpha=mu*(1-rho)/rho;
    beta=(1-mu)*(1-rho)/rho;

    ll= lgamma(n+1)+lgamma(success+alpha)+lgamma(n-success+beta)+lgamma(alpha+beta)
      -lgamma(success+1)-lgamma(n-success+1)-lgamma(n+alpha+beta) -lgamma(alpha)-lgamma(beta);
    model success ~ general(ll);

   * Estimate the treatment effect;
    estimate "LogOR" b_treat;
    ods output AdditionalEstimates=LogOR_BBST__temp(where=(Label="LogOR") rename=(Estimate=LogOR_BBST_ StandardError=SE_LogOR_BBST_));
    title"Beta-binomial model, common-rho";
run;
ods select all;


  * Confidence intervals with k-1 degrees of freedom;
  data LogOR_BBST_(keep=simruns LogOR_BBST_ CI95L_LogOR_BBST_ CI95U_LogOR_BBST_);
    set LogOR_BBST__temp;
    CI95L_LogOR_BBST_ = LogOR_BBST_ - quantile('T',.975,&nstudy*2-2)* SE_LogOR_BBST_;
    CI95U_LogOR_BBST_ = LogOR_BBST_ + quantile('T',.975,&nstudy*2-2)* SE_LogOR_BBST_;
run;
```

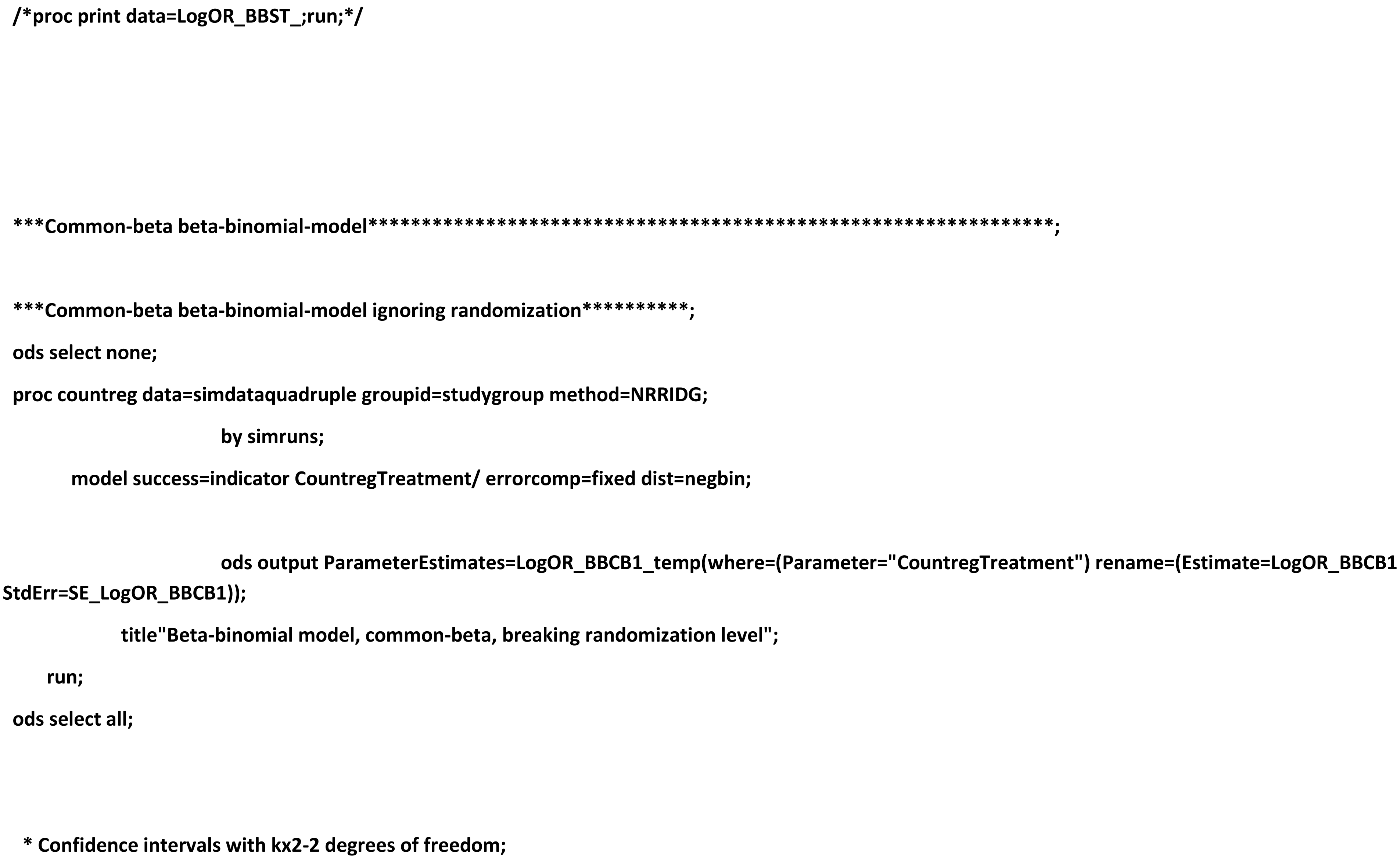

```
/*proc print data=LogOR_BBST_;run;*/




***Common-beta beta-binomial-model****************************************************************;


***Common-beta beta-binomial-model ignoring randomization**********;

ods select none;

proc countreg data=simdataquadruple groupid=studygroup method=NRRIDG;

                    by simruns;

      model success=indicator CountregTreatment/ errorcomp=fixed dist=negbin;


                    ods output ParameterEstimates=LogOR_BBCB1_temp(where=(Parameter="CountregTreatment") rename=(Estimate=LogOR_BBCB1
StdErr=SE_LogOR_BBCB1));

           title"Beta-binomial model, common-beta, breaking randomization level";

    run;

ods select all;


 * Confidence intervals with kx2-2 degrees of freedom;
```

```
data LogOR_BBCB1(keep=simruns LogOR_BBCB1 CI95L_LogOR_BBCB1 CI95U_LogOR_BBCB1);
  set LogOR_BBCB1_temp;
  CI95L_LogOR_BBCB1 = LogOR_BBCB1 - quantile('T',.975,&nstudy*2-2)* SE_LogOR_BBCB1;
  CI95U_LogOR_BBCB1 = LogOR_BBCB1 + quantile('T',.975,&nstudy*2-2)* SE_LogOR_BBCB1;
run;
/*proc print data=LogOR_BBCB1;run;*/



***Common-beta beta-binomial-model preserving randomization**********;
ods select none;
proc countreg data=simdataquadruple groupid=study method=NRRIDG;
                              by simruns;
        model success=indicator CountregTreatment/ errorcomp=fixed dist=negbin;


                              ods output ParameterEstimates=LogOR_BBCB2_temp(where=(Parameter="CountregTreatment") rename=(Estimate=LogOR_BBCB2
StdErr=SE_LogOR_BBCB2));
                    title"Beta-binomial model, common-beta with overdispersion term";
    run;
ods select all;
```

```
 * Confidence intervals with kx2-2 degrees of freedom;
 data LogOR_BBCB2(keep=simruns LogOR_BBCB2 CI95L_LogOR_BBCB2 CI95U_LogOR_BBCB2);
   set LogOR_BBCB2_temp;
   CI95L_LogOR_BBCB2 = LogOR_BBCB2 - quantile('T',.975,&nstudy*2-2)* SE_LogOR_BBCB2;
   CI95U_LogOR_BBCB2 = LogOR_BBCB2 + quantile('T',.975,&nstudy*2-2)* SE_LogOR_BBCB2;
 run;
 /*proc print data=LogOR_BBCB2;run;*/




************************************************************************************************************************;
*** GLMM models ******************************************************************************************************;
************************************************************************************************************************;


 ***GLMM with random intercept and random treatment effect***************************************;
 ods select none;
 proc glimmix data=simdatadouble order=formatted method=quad(qpoints=1);
    by simruns;
```

```
    class treatment;
    model success/n= treatment / d=bin link=logit solution cl;
    random intercept treatment / subject=study type=vc;
        nloptions tech=NRRIDG;

    ods output ParameterEstimates=LogOR_GLRR_temp(where=(treatment=1)
                    rename=(Estimate=LogOR_GLRR_ StdErr=SE_LogOR_GLRR_
                    Lower=CI95L_LogOR_GLRR_ Upper=CI95U_LogOR_GLRR_));
        format treatment treatf.;
        title"GLMM, random study effect and random treatment effect";
run;
ods select all;

data LogOR_GLRR_(keep=simruns LogOR_GLRR_ CI95L_LogOR_GLRR_ CI95U_LogOR_GLRR_);
        set LogOR_GLRR_temp;
run;
/*proc print data=LogOR_GLRR_;run;*/
```

```
*********************************************************************************************************************;
*** Inverse variance models **************************************************************************************;
*********************************************************************************************************************;

****Knapp-Hartung-Verfahren mit Paule-Mandel Heterogenit tssch tzer************************************;

  *prepare dataset for Paule and Mandel Macro;
  data pmivre;
    set simdatacontinuity;
    vari=1/weight_test_LogOR;
  run;

   * Determine the number of studies;
          ods select none;
          proc means data=pmivre n;
        var study;
        by simruns;
        output out=outmeans n(study)=nstudies;
   run;
```

```
  ods select all;


  * ... and merge with the original data set;
   data pmstart (keep=simruns nstudies study test_LogOR vari);
       merge outmeans pmivre;
       by simruns;
   run;


*macro to estimate the Paule and Mandel heterogenity variance etimator Tau(pm);
%macro PmEstimator(data=, start_tau_square_pm=, nstudies=, iterations=);
   data pm;
   set &data;
          by simruns;
   if tau_square_pm=. then tau_square_pm=&start_tau_square_pm;


     %let max_fin_f_tau=.;
     %let _i=0;


     %put &max_fin_f_tau;
     %put &_i;
```

```
 *starting the loop;
 %do %until(&max_fin_f_tau<=0 or &_i=&iterations);
 %let _i = %eval(&_i+1);


  run;
/*proc print data=pm;*/



 *calculate weigths and terms that are necessary to calculate the Paule and Mandel estimating equation;
   data pm;
     set pm;
     by simruns;

     * Initialize terms that are summed up;
     if first.simruns then do;
       sum_w_pm=0; sum_num_yw_tau=0;
     end;

   * Calculate terms that are summed across studies ...;
     w_pm=1/(tau_square_pm+vari);
     num_yw_tau=w_pm*test_LogOR;
```

```
    sum_w_pm+w_pm;
    sum_num_yw_tau+num_yw_tau;

   * and sum them...;

    if last.simruns then do;
       yw_tau=sum_num_yw_tau/sum_w_pm;
    end;
   run;

*imput missing column values necessary for further summed terms...;
 data pm;
   do until(yw_tau);
   set pm end=last;
   if not missing(yw_tau) then complet_yw_tau=yw_tau;
   end;

   do until(yw_tau);
   set pm end=_last;
```

```
   output;
   end;
 run;

/*proc print data=pm;run;*/

 *calulate the Paule and Mandel estimating equation (f_tau), the target equation (fin_f_tau) and the correction (delta_tau);
data pm;
set pm;
by simruns;
   * Initialize terms that are summed up;
   if first.simruns then do;
     sum_f_tau=0;
     sum_deno_delta_tau=0;
   end;

     f_tau=w_pm*(test_LogOR-complet_yw_tau)**2;
     deno_delta_tau=w_pm**2*(test_LogOR-complet_yw_tau)**2;

   * Calculate terms that are summed across studies ...;
     sum_f_tau+f_tau;
```

```
    sum_deno_delta_tau+deno_delta_tau;

   * and sum them...;
   if last.simruns then do;
    a_f_tau=sum_f_tau-(&nstudies-1);
    fin_f_tau=max(a_f_tau,0);

    delta_tau=max(0, (sum_f_tau-(&nstudies-1))/sum_deno_delta_tau);

   end;
run;



   *determine the maximum of fin_f_tau across meta-analysis so that the loop is repeated until the stoping thresehold is reached in all meta-analysis;
          ods select none;
   proc means data=pm n;
    var study;
    output out=outmeanspm max(fin_f_tau)=max_fin_f_tau;

   run;
          ods select all;
```

```
*assign the actuall max_fin_f_tau as macro variable;
data pmeins;
set outmeanspm;
   call symputx('max_fin_f_tau', max_fin_f_tau);
run;

   %put &max_fin_f_tau;
run;

data pm;
set pm;
by simruns;
  * Initialize terms that are summed up;
  if first.simruns then do;
    sum_tau_square_pm_temp1=0;
  end;

  tau_square_pm_temp1=tau_square_pm+delta_tau;

  * Calculate terms that are summed across studies ...;
```

```
   sum_tau_square_pm_temp1+tau_square_pm_temp1;


         * and sum them...;
   if last.simruns then do;
     tau_square_pm_temp1=sum_tau_square_pm_temp1;
   end;
run;

*imput missing column values of tau_square_pm which are necessary for summed terms in the next loop;
data pm;
   do until(tau_square_pm_temp1 or last);
   set pm end=last;
   if not missing(tau_square_pm_temp1) then  tau_square_pm_neu=tau_square_pm_temp1;
   end;

   do until(tau_square_pm_temp1 or _last);
   set pm end=_last;

   output;
   end;
run;
```

```
*drop variables that depend on tau_square_pm and conseqeuently change values in each loop;
data pm;
set pm;
by simruns;
tau_square_pm=tau_square_pm_neu;
drop yw_tau sum_w_pm sum_num_yw_tau w_pm num_yw_tau complet_yw_tau sum_f_tau sum_deno_delta_tau f_tau deno_delta_tau a_f_tau
fin_f_tau delta_tau sum_tau_square_pm_temp1 tau_square_pm_temp1 tau_square_pm_neu;
run;

/*proc print data=pm;run;*/

%end;

%mend PmEstimator;


%PmEstimator(data=pmstart, start_tau_square_pm=0.00001, nstudies=nstudies, iterations=100);
```

```
*prepare data for the Paule and Mandel based REM;
 data rempmeins;
      set pm;
                    by simruns;
     * Compute the random effects weight for each study;
      rem_w_pm=1/(vari+tau_square_pm);
      run;



* Compute the random effects estimator theta_rem and its variance;
 data rempm(keep=simruns nstudies LogOR_IVRE_PM rem_w_pm var_theta_rem_pm se_theta_rem_pm tau_square_pm);
      set rempmeins end=lastrecord;
      by simruns;

     * Initialize terms that are summed up;
      if first.simruns then do;
      sum_rem_times_w_pm=0;sum_rem_w_pm=0;
      end;

     * Calculate terms that are summed across studies ...;
      test_LogOR_times_rem_w_pm = rem_w_pm*test_LogOR;
```

```
    * ... and sum them;
     sum_rem_w_pm+rem_w_pm;
     sum_rem_times_w_pm + test_LogOR_times_rem_w_pm;

    * When the last observation is reached, calculate estimates and output them;
     if last.simruns then do;
     LogOR_IVRE_PM=sum_rem_times_w_pm/sum_rem_w_pm;
     var_theta_rem_pm=1/sum_rem_w_pm;
     se_theta_rem_pm=1/sqrt(sum_rem_w_pm);

     output;
    end;

     title"rempm";
run;

* ... explode the data set rempm...;
data rempmexplode; set rempm; by simruns; do study=1 to nstudies; output; end; run;

* ... and merge with the data set rempmeins;
```

```
data pmfinal;
   merge rempmeins(keep=simruns study test_LogOR rem_w_pm) rempmexplode(keep=simruns study nstudies LogOR_IVRE_PM var_theta_rem_pm
                                                       se_theta_rem_pm tau_square_pm);
   by simruns study;
run;
/*proc print data=pmfinal;run;*/



*Compute the confidence interval adjustment term q;
data LogOR_HKPMtemp(keep=simruns nstudies LogOR_IVRE_PM se_theta_rem_pm tau_square_pm q);
   set pmfinal end=lastrecord;
   by simruns;


 * Initialize terms that are summed up;
   if first.simruns then do;
     sum_num_HKPM_=0;
   end;

 * Calculate Terms that are summed across studies ...;
```

```
num_HKPM_ = rem_w_pm*(test_LogOR-LogOR_IVRE_PM)**2;


* ... and sum them;
sum_num_HKPM_+num_HKPM_;



* When the last observation is reached calculate estimates and output them;
if last.simruns then do;
q=sum_num_HKPM_/(nstudies-1);
   output;
  end;


      run;




****Knapp-Hartung-Verfahren orginal***************************************************************************;


*calculate confidence intervals;
data LogOR_HKPM_(keep=simruns nstudies LogOR_HKPM_ CI95L_LogOR_HKPM_ CI95U_LogOR_HKPM_ tau_square_pm q);
   set LogOR_HKPMtemp;
```

```
LogOR_HKPM_= LogOR_IVRE_PM;
by simruns;

CI95L_LogOR_HKPM_ =  LogOR_HKPM_ - quantile('T',.975,nstudies-1)* sqrt(q)*se_theta_rem_pm;
CI95U_LogOR_HKPM_ =  LogOR_HKPM_ + quantile('T',.975,nstudies-1)* sqrt(q)*se_theta_rem_pm;

run;
/*proc print data=LogOR_HKPM_;title"Hartung/Knapp (PM) ohne Varianzkorrektur"; run;*/


****Knapp-Hartung-Verfahren modifiziert**********************************************************************;
data LogOR_MHKPM(keep=simruns nstudies LogOR_MHKPM CI95L_LogOR_MHKPM CI95U_LogOR_MHKPM);
  set LogOR_HKPMtemp;
      LogOR_MHKPM=LogOR_IVRE_PM;
  by simruns;

  CI95L_LogOR_MHKPM =  LogOR_MHKPM - quantile('T',.975,nstudies-1)* se_theta_rem_pm;
  CI95U_LogOR_MHKPM =  LogOR_MHKPM + quantile('T',.975,nstudies-1)* se_theta_rem_pm;
```

```
run;
/*proc print data=LogOR_MHKPM;title"Hartung/Knapp (PM) mit Varianzkorrektur"; run;*/



***************************************************************************************************************************************************
*******************************************;
***************************************************************************************************************************************************
*******************************************;
********** SUMMARIZE SIMULATION RESULTS
***************************************************************************************************************************************************
***;
***************************************************************************************************************************************************
*******************************************;
***************************************************************************************************************************************************
*******************************************;

data result_&ident;


  * Collect information for this simulation run;
   Ident="&ident";
```

```
* Merge result data sets from the different procedures;
  merge simdata(keep=simruns true_pc true_LogOR true_OR true_TauSquare study nt st nc sc pt pc)
                  simdatacontinuity(keep=simruns test_OR weight_test_LogOR)
                  LogOR_BBST_ LogOR_BBCB1 LogOR_BBCB2
                  LogOR_GLRR_ LogOR_HKPM_ LogOR_MHKPM
                  ;
  by simruns;

* Macro to calculate performance measures;
  %macro calcresp(method,true,nullvalue);
                  bias_&method=&method-true_&true;
                  mse_&method=(&method-true_&true)**2;
     coverage_&method=.;
     if CI95L_&method<true_&true and CI95U_&method>true_&true then coverage_&method=1;
     if CI95L_&method>true_&true or  CI95U_&method<true_&true then coverage_&method=0;
     if &method=. or CI95L_&method=. or CI95U_&method=. then coverage_&method=.;
                  ci_length_&method=abs(CI95U_&method-CI95L_&method);
     power_&method=.;
     if CI95L_&method>&nullvalue or  CI95U_&method<&nullvalue then power_&method=1;
     if CI95L_&method<&nullvalue and CI95U_&method>&nullvalue then power_&method=0;
```

```
      if &method=. or CI95L_&method=. or CI95U_&method=. then power_&method=.;
   %mend calcresp;

   %calcresp(LogOR_BBST_,LogOR,0);
      %calcresp(LogOR_BBCB1,LogOR,0);
      %calcresp(LogOR_BBCB2,LogOR,0);
      %calcresp(LogOR_GLRR_,LogOR,0);
      %calcresp(LogOR_HKPM_,LogOR,0);
      %calcresp(LogOR_MHKPM,LogOR,0);

 run;

*Delete data sets;

 proc datasets;
   delete Simdata ident simdatadouble simdatahelp simdataquadruple simdataexplode simdatadouble_temp1 simdatacontinuity SimpleStats_pt
SimpleStats_pc SimpleStats_OR
                 startingvalues_Beta_Binomial LogOR_BBST__temp LogOR_BBST_ LogOR_BBCB1_temp LogOR_BBCB1 LogOR_BBCB2_temp
LogOR_BBCB2
                 LogOR_GLRR_temp LogOR_GLRR_ outmeans outmeanspm logor_hkpmtemp pmivre  pmstart
```

```
                    pm pmeins rempmeins rempm rempmexplode pmfinal LogOR_HKPM_ LogOR_MHKPM;

  quit;

%mend RegressionModelsSim;


***************************************************************************************************************************************************
*******************************************;

***************************************************************************************************************************************************
*******************************************;

********** RUN THE SIMULATION
***************************************************************************************************************************************************
**************;

***************************************************************************************************************************************************
*******************************************;

***************************************************************************************************************************************************
*******************************************;


 /*options mlogic mprint;*/


 %RegressionModelsSim(nstudy=1,  baselineprob=0.05,  effect=1,  rem=1,   seedsim=02, nsimruns=5000, printsimdata=0);
```

```
/*proc print data=result_h1remnc; run; title "simulation results";*/

libname results "C:\Users\mathes1\Desktop\SAS";
data results.result_h105remNC;
set result_h105remNC;
run;
```

**Supplemental figures**

Figure S2: converged runs under H1

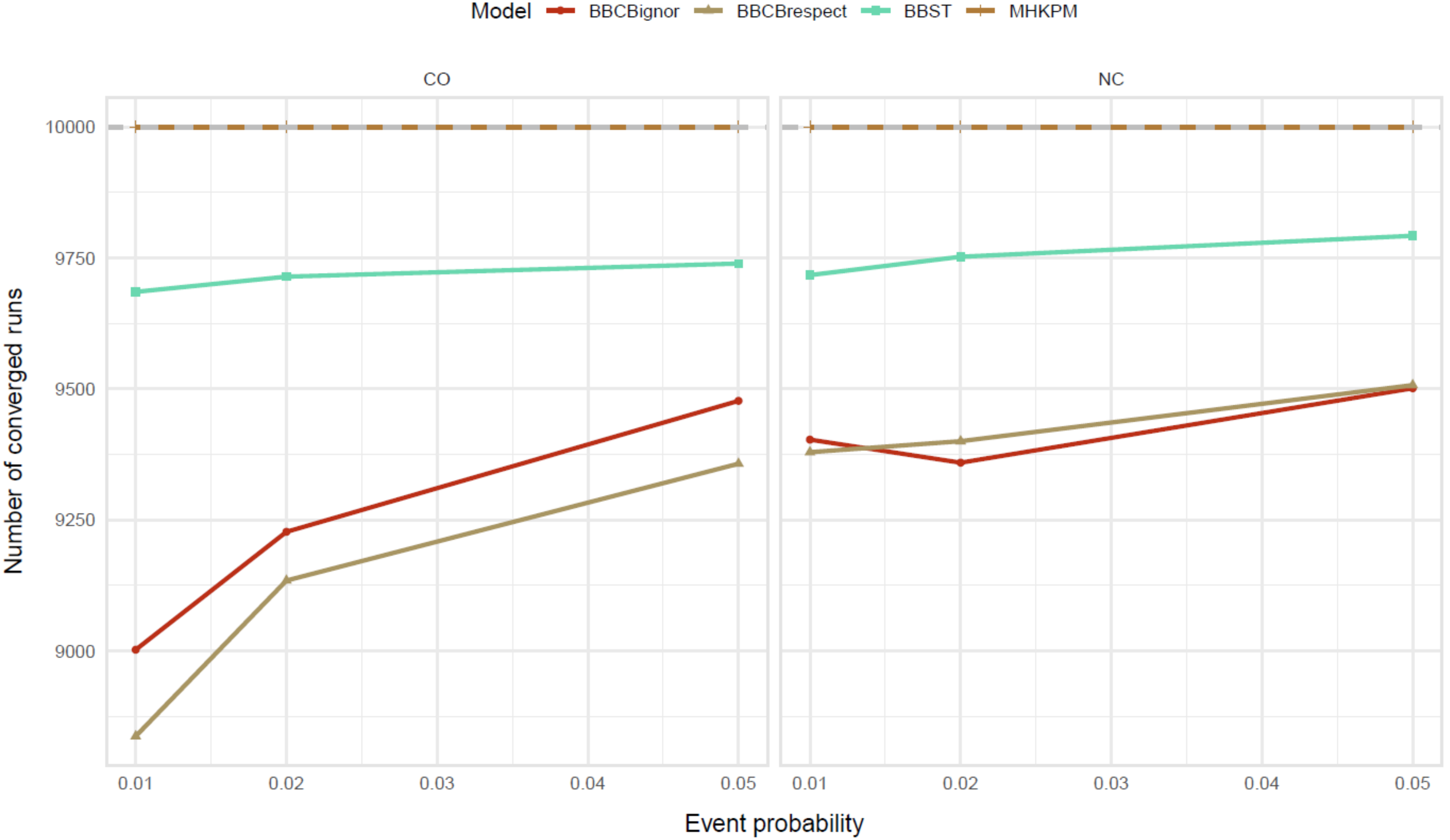

Figure S2: Bias under H1

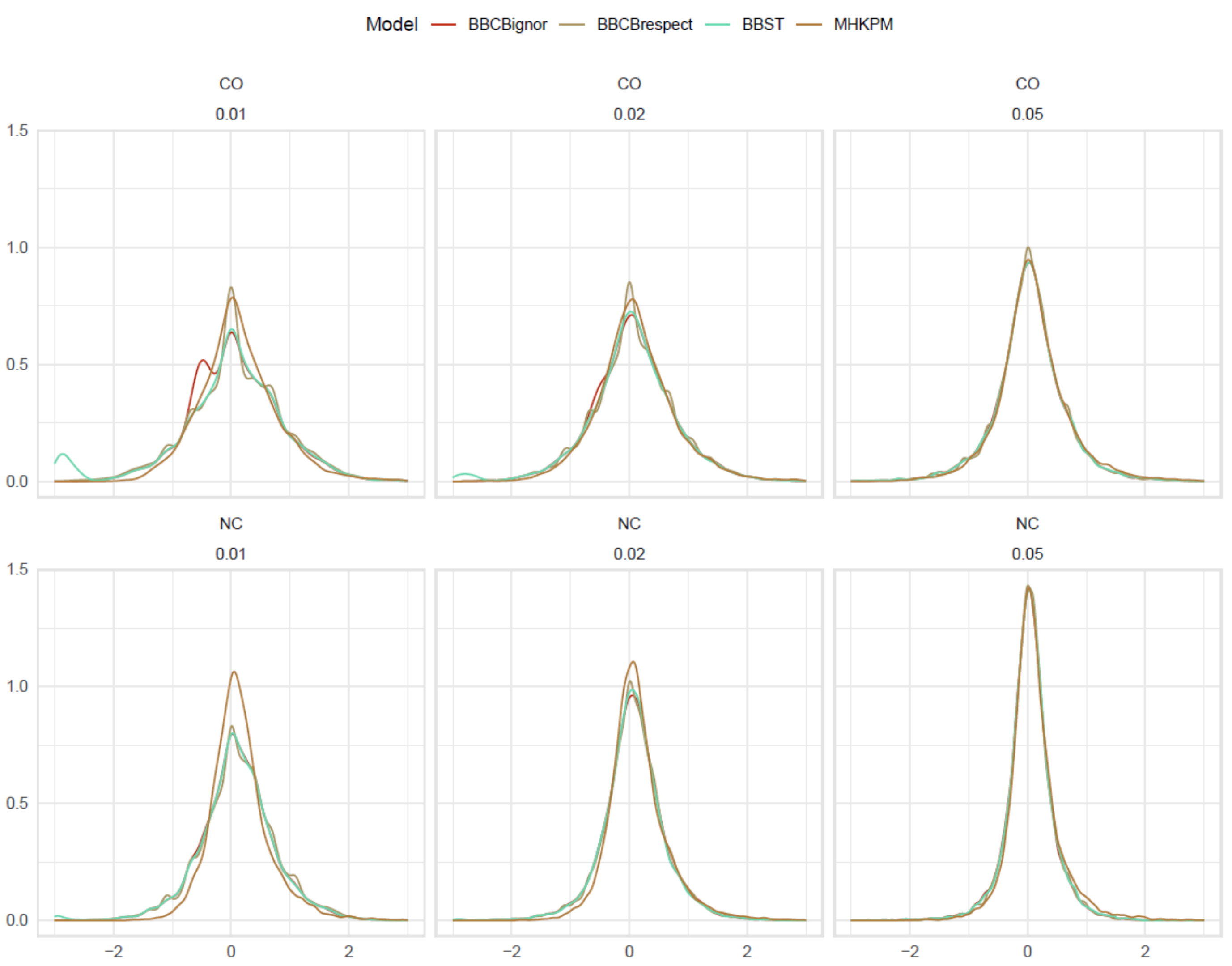

Figure S3: Coverage probability under H1

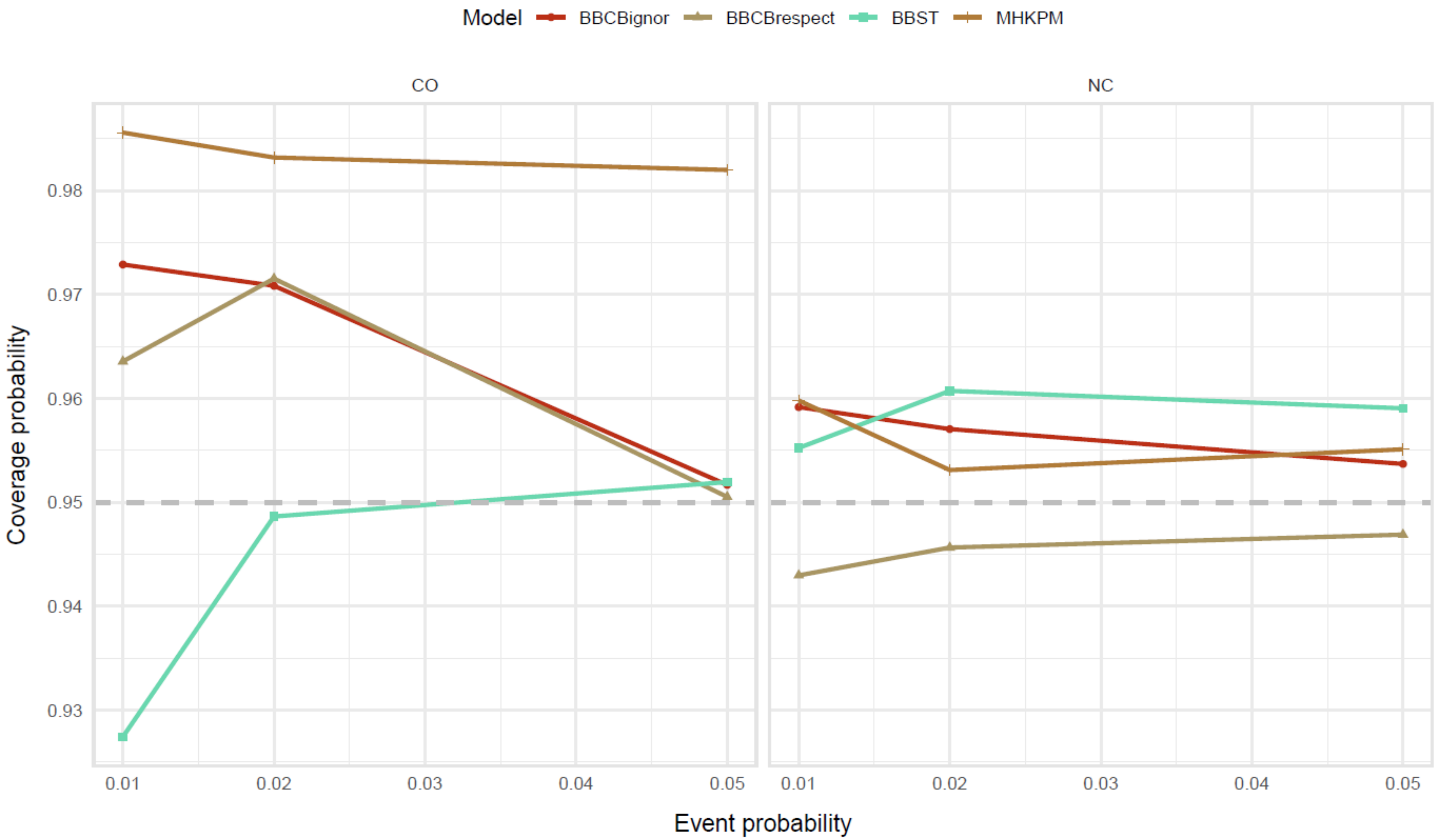

Figure S4: length of the 95%-confidence intervals under H1

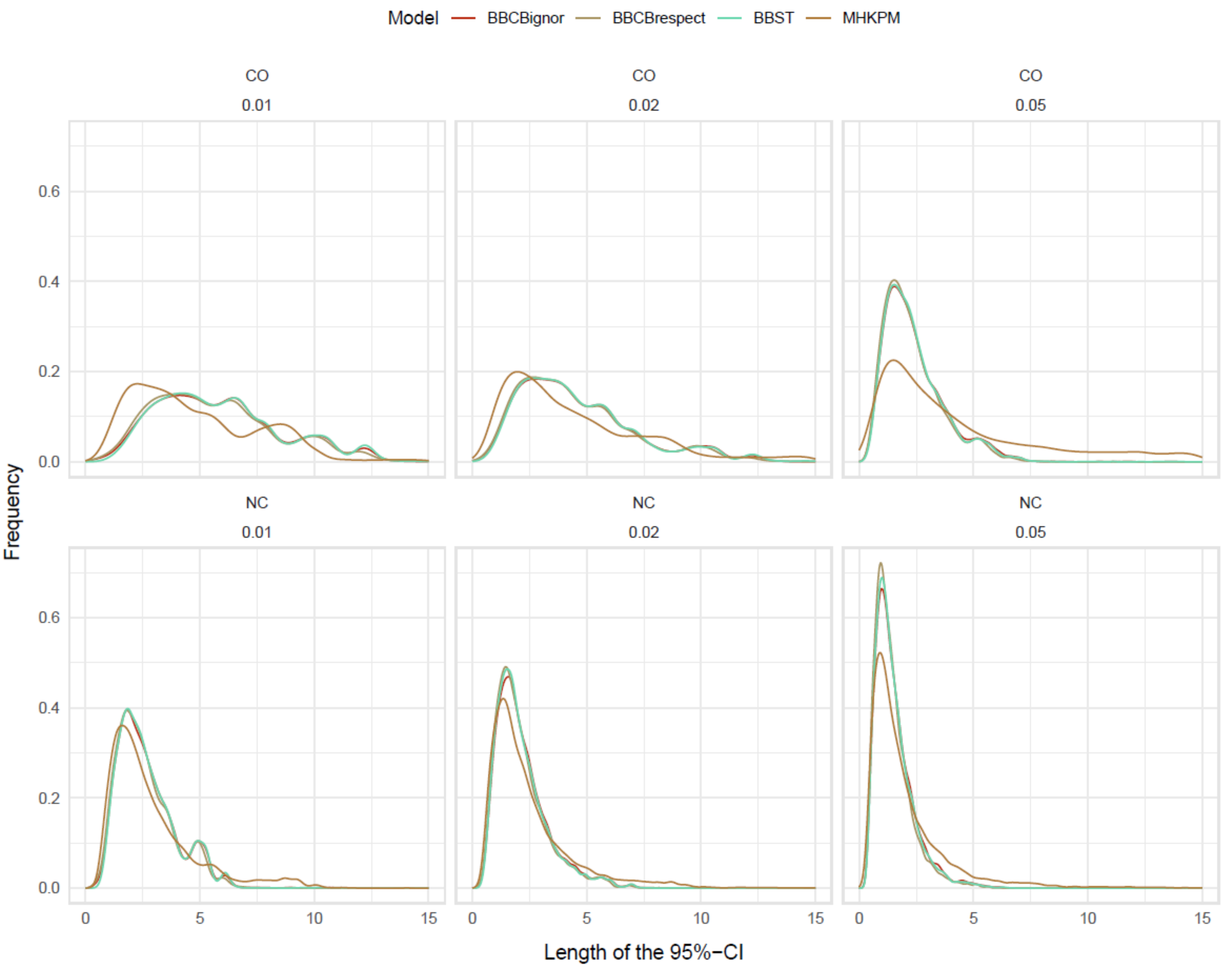

**Example meta-analysis**

| | Individual study results | | | | | Meta-analysis result | | | |
|---|---|---|---|---|---|---|---|---|---|
| | Study | n treatment | n control | Events Treatment | Events control | Tau square | LogOR BBCB ignor | LogOR BBCB respect | LogOR MKHPM |
| Example different study sizes | | | | | | | | | |
| | 1 | 11 | 10 | 0 | 0 | 0.00001 | 4,04312E+14 | -0.2877 | -0.2655 |
| | 2 | 23 | 16 | 1 | 2 | | | | |
| | 3 | 10 | 14 | 0 | 0 | | | | |
| | 4 | 55 | 55 | 0 | 0 | | | | |
| | 5 | 293 | 321 | 5 | 5 | | | | |
| | 6 | 12 | 11 | 0 | 0 | | | | |
| | 7 | 36 | 34 | 0 | 1 | | | | |
| Example heterogenity | | | | | | | | | |
| | 1 | 255 | 255 | 9 | 3 | 1,51037E+14 | -0.7100 | 7,45498E+13 | -0.3160 |
| | 2 | 166 | 160 | 0 | 3 | | | | |
| | 3 | 98 | 86 | 0 | 1 | | | | |

**Declaration of conflicting interest**

The authors declare that they have no financial competing interests. OK and TM have a scientific conflict of interest as OK suggested using the BB model for meta-analysis and developed the suggested extensions and were involved in various previous simulation studies on BB models.

**Funding**

The study received no funding.

**Authors' contributions**

TM: idea, development of methods, design of study and simulation, programming and performing simulation, interpretation of results, writing manuscript

MS: visualization, revision of manuscript

OK: idea, development of methods, design of study and simulation, interpretation of results, revision of manuscript

## References

1. Schulz M, Kramer M, Kuss O, et al. A re-analysis of about 60,000 sparse data meta-analyses suggests that using an adequate method for pooling matters. *Research Synthesis Methods* 2024; 15: 978-987. DOI: https://doi.org/10.1002/jrsm.1748.
2. Xu C, Zhou X, Zorzela L, et al. Utilization of the evidence from studies with no events in meta-analyses of adverse events: an empirical investigation. *BMC Medicine* 2021; 19: 141. DOI: 10.1186/s12916-021-02008-2.
3. Zabriskie BN, Cole N, Baldauf J, et al. The impact of correction methods on rare-event meta-analysis. *Research Synthesis Methods* 2024; 15: 130-151. DOI: https://doi.org/10.1002/jrsm.1677.
4. Bhaumik DK, Amatya A, Normand SL, et al. Meta-Analysis of Rare Binary Adverse Event Data. *J Am Stat Assoc* 2012; 107: 555-567. DOI: 10.1080/01621459.2012.664484.
5. Günhan BK, Röver C and Friede T. Random-effects meta-analysis of few studies involving rare events. *Research Synthesis Methods* 2020; 11: 74-90. DOI: https://doi.org/10.1002/jrsm.1370.
6. Jansen K and Holling H. Rare events meta-analysis using the Bayesian beta-binomial model. *Research Synthesis Methods* 2023; 14: 853-873. DOI: https://doi.org/10.1002/jrsm.1662.
7. Kuss O. Statistical methods for meta-analyses including information from studies without any events-add nothing to nothing and succeed nevertheless. *Stat Med* 2015; 34: 1097-1116. 20141201. DOI: 10.1002/sim.6383.
8. Simmonds MC and Higgins JPT. A general framework for the use of logistic regression models in meta-analysis. *Statistical Methods in Medical Research* 2014; 25: 2858-2877. DOI: 10.1177/0962280214534409.
9. Beisemann M, Doebler P and Holling H. Comparison of random-effects meta-analysis models for the relative risk in the case of rare events: A simulation study. *Biometrical Journal* 2020; 62: 1597-1630. DOI: https://doi.org/10.1002/bimj.201900379.
10. Felsch M, Beckmann L, Bender R, et al. Performance of several types of beta-binomial models in comparison to standard approaches for meta-analyses with very few studies. *BMC Medical Research Methodology* 2022; 22: 319. DOI: 10.1186/s12874-022-01779-3.

11. Jansen K and Holling H. Random-effects meta-analysis models for the odds ratio in the case of rare events under different data-generating models: A simulation study. *Biometrical Journal* 2023; 65: 2200132. DOI: https://doi.org/10.1002/bimj.202200132.
12. Mathes T and Kuss O. A comparison of methods for meta-analysis of a small number of studies with binary outcomes. *Research Synthesis Methods* 2018; 9: 366-381. DOI: https://doi.org/10.1002/jrsm.1296.
13. Karahalios A, McKenzie JE and White IR. Contrast-Based and Arm-Based Models for Network Meta-Analysis. *Methods Mol Biol* 2022; 2345: 203-221. DOI: 10.1007/978-1-0716-1566-9_13.
14. Mathes T and Kuss O. Beta-binomial models for meta-analysis with binary outcomes: Variations, extensions, and additional insights from econometrics. *Research Methods in Medicine & Health Sciences* 2021; 2: 82-89. DOI: 10.1177/2632084321996225.
15. Hausman J, Hall BH and Griliches Z. Econometric Models for Count Data with an Application to the Patents-R & D Relationship. *Econometrica* 1984; 52: 909-938. DOI: 10.2307/1911191.
16. Guimarães P. A Simple Approach to Fit the Beta-binomial Model. *The Stata Journal* 2005; 5: 385-394. DOI: 10.1177/1536867x0500500307.
17. Page MJ, Shamseer L, Altman DG, et al. Epidemiology and Reporting Characteristics of Systematic Reviews of Biomedical Research: A Cross-Sectional Study. *PLOS Medicine* 2016; 13: e1002028. DOI: 10.1371/journal.pmed.1002028.
18. Turner RM, Davey J, Clarke MJ, et al. Predicting the extent of heterogeneity in meta-analysis, using empirical data from the Cochrane Database of Systematic Reviews. *Int J Epidemiol* 2012; 41: 818-827. 20120329. DOI: 10.1093/ije/dys041.
19. Deeks JJ, Higgins JP, Altman DG, et al. Analysing data and undertaking meta-analyses. *Cochrane Handbook for Systematic Reviews of Interventions*. 2019, pp.241-284.
20. Knapp G and Hartung J. Improved tests for a random effects meta-regression with a single covariate. *Stat Med* 2003; 22: 2693-2710. Journal Article 2003/08/27. DOI: 10.1002/sim.1482.
21. Møller CH, Penninga L, Wetterslev J, et al. Off-pump versus on-pump coronary artery bypass grafting for ischaemic heart disease. *Cochrane Database of Systematic Reviews* 2012. DOI: 10.1002/14651858.CD007224.pub2.
22. Rücker G and Schumacher M. Simpson's paradox visualized: the example of the rosiglitazone meta-analysis. *BMC Med Res Methodol* 2008; 8: 34. 20080530. DOI: 10.1186/1471-2288-8-34.
23. Piepho H-P and Madden LV. How to observe the principle of concurrent control in an arm-based meta-analysis using SAS procedures GLIMMIX and BGLIMM. *Research Synthesis Methods* 2022; 13: 821-828. DOI: https://doi.org/10.1002/jrsm.1576.
24. Jackson D, Law M, Stijnen T, et al. A comparison of seven random-effects models for meta-analyses that estimate the summary odds ratio. *Stat Med* 2018; 37: 1059-1085. 20180108. DOI: 10.1002/sim.7588.